\documentclass[twocolumn]{aastex631}
\usepackage{graphicx}	
\usepackage{amsmath}	

\usepackage{longtable}

\begin{document}

\title[]{GECAM Observations of the Galactic Magnetar SGR J1935+2154 during the 2021 and 2022 Burst Active Episodes. I. Burst Catalog}
\shortauthors{Xie et al.}

\author[0000-0001-9217-7070]{Sheng-Lun Xie}
\affiliation{Institute of Astrophysics, Central China Normal University, Wuhan 430079, China}
\affiliation{Key Laboratory of Particle Astrophysics, Institute of High Energy Physics, Chinese Academy of Sciences, 19B Yuquan Road, Beijing 100049, China}
%
\author[0000-0002-6540-2372]{Ce Cai}
\affiliation{College of Physics and Hebei Key Laboratory of Photophysics Research and Application, Hebei Normal University, Shijiazhuang, Hebei 050024, China}
\correspondingauthor{Ce Cai}
\email{caice@hebtu.edu.cn}
\author[0000-0002-1067-1911]{Yun-Wei Yu}
\affiliation{Institute of Astrophysics, Central China Normal University, Wuhan 430079, China}
\correspondingauthor{Yun-Wei Yu}
\email{yuyw@ccnu.edu.cn}
\author[0000-0002-4771-7653]{Shao-Lin Xiong}
\affiliation{Key Laboratory of Particle Astrophysics, Institute of High Energy Physics, Chinese Academy of Sciences, 19B Yuquan Road, Beijing 100049, China}
\correspondingauthor{Shao-Lin Xiong}
\email{xiongsl@ihep.ac.cn}
\author[0000-0002-0633-5325]{Lin Lin}
\affiliation{Department of Astronomy, Beijing Normal University, Beijing 100088, China}
%
\author[0000-0002-4636-0293]{Yi Zhao}
\affiliation{School of Computer and Information, Dezhou University, Dezhou 253023, China}
%
\author[0000-0001-5586-1017]{Shuang-Nan Zhang}
\affiliation{Key Laboratory of Particle Astrophysics, Institute of High Energy Physics, Chinese Academy of Sciences, 19B Yuquan Road, Beijing 100049, China}
%
\author[0000-0003-0274-3396]{Li-Ming Song}
\affiliation{Key Laboratory of Particle Astrophysics, Institute of High Energy Physics, Chinese Academy of Sciences, 19B Yuquan Road, Beijing 100049, China}
%
\author{Ping Wang}
\affiliation{Key Laboratory of Particle Astrophysics, Institute of High Energy Physics, Chinese Academy of Sciences, 19B Yuquan Road, Beijing 100049, China}
%
\author{Xiao-Bo Li}
\affiliation{Key Laboratory of Particle Astrophysics, Institute of High Energy Physics, Chinese Academy of Sciences, 19B Yuquan Road, Beijing 100049, China}
%
\author{Wang-Chen Xue}
\affiliation{Key Laboratory of Particle Astrophysics, Institute of High Energy Physics, Chinese Academy of Sciences, 19B Yuquan Road, Beijing 100049, China}
\affiliation{University of Chinese Academy of Sciences, Beijing 100049, China}
%
\author{Peng Zhang}
\affiliation{College of Electronic and Information Engineering, Tongji University, Shanghai 201804, China}
\affiliation{Key Laboratory of Particle Astrophysics, Institute of High Energy Physics, Chinese Academy of Sciences, 19B Yuquan Road, Beijing 100049, China}
%
\author{Chao Zheng}
\affiliation{Key Laboratory of Particle Astrophysics, Institute of High Energy Physics, Chinese Academy of Sciences, 19B Yuquan Road, Beijing 100049, China}
\affiliation{University of Chinese Academy of Sciences, Beijing 100049, China}
%
\author{Yan-Qiu Zhang}
\affiliation{Key Laboratory of Particle Astrophysics, Institute of High Energy Physics, Chinese Academy of Sciences, 19B Yuquan Road, Beijing 100049, China}
\affiliation{University of Chinese Academy of Sciences, Beijing 100049, China}
%
\author{Jia-Cong Liu}
\affiliation{Key Laboratory of Particle Astrophysics, Institute of High Energy Physics, Chinese Academy of Sciences, 19B Yuquan Road, Beijing 100049, China}
\affiliation{University of Chinese Academy of Sciences, Beijing 100049, China}
%
\author{Chen-Wei Wang}
\affiliation{Key Laboratory of Particle Astrophysics, Institute of High Energy Physics, Chinese Academy of Sciences, 19B Yuquan Road, Beijing 100049, China}
\affiliation{University of Chinese Academy of Sciences, Beijing 100049, China}
%
\author{Wen-Jun Tan}
\affiliation{Key Laboratory of Particle Astrophysics, Institute of High Energy Physics, Chinese Academy of Sciences, 19B Yuquan Road, Beijing 100049, China}
\affiliation{University of Chinese Academy of Sciences, Beijing 100049, China}
%
\author{Yue Wang}
\affiliation{Key Laboratory of Particle Astrophysics, Institute of High Energy Physics, Chinese Academy of Sciences, 19B Yuquan Road, Beijing 100049, China}
\affiliation{University of Chinese Academy of Sciences, Beijing 100049, China}
%
\author{Zheng-Hang Yu}
\affiliation{Key Laboratory of Particle Astrophysics, Institute of High Energy Physics, Chinese Academy of Sciences, 19B Yuquan Road, Beijing 100049, China}
\affiliation{University of Chinese Academy of Sciences, Beijing 100049, China}
%
\author{Pei-Yi Feng}
\affiliation{Key Laboratory of Particle Astrophysics, Institute of High Energy Physics, Chinese Academy of Sciences, 19B Yuquan Road, Beijing 100049, China}
\affiliation{University of Chinese Academy of Sciences, Beijing 100049, China}
%
\author{Jin-Peng Zhang}
\affiliation{Key Laboratory of Particle Astrophysics, Institute of High Energy Physics, Chinese Academy of Sciences, 19B Yuquan Road, Beijing 100049, China}
\affiliation{University of Chinese Academy of Sciences, Beijing 100049, China}
%
\author{Shuo Xiao}
\affiliation{Guizhou Provincial Key Laboratory of Radio Astronomy and Data Processing, Guizhou Normal University, Guiyang 550001, China}
%
\author{Hai-Sheng Zhao}
\affiliation{Key Laboratory of Particle Astrophysics, Institute of High Energy Physics, Chinese Academy of Sciences, 19B Yuquan Road, Beijing 100049, China}
%
\author{Wen-Long Zhang}
\affiliation{School of Physics and Physical Engineering, Qufu Normal University, Qufu, Shandong 273165, China}
\affiliation{Key Laboratory of Particle Astrophysics, Institute of High Energy Physics, Chinese Academy of Sciences, 19B Yuquan Road, Beijing 100049, China}
%
\author{Yan-Ting Zhang}
\affiliation{Key Laboratory of Particle Astrophysics, Institute of High Energy Physics, Chinese Academy of Sciences, 19B Yuquan Road, Beijing 100049, China}
%
\author{Yue Huang}
\affiliation{Key Laboratory of Particle Astrophysics, Institute of High Energy Physics, Chinese Academy of Sciences, 19B Yuquan Road, Beijing 100049, China}
%
\author{Xiao-Yun Zhao}
\affiliation{Key Laboratory of Particle Astrophysics, Institute of High Energy Physics, Chinese Academy of Sciences, 19B Yuquan Road, Beijing 100049, China}
%
\author{Xiang Ma}
\affiliation{Key Laboratory of Particle Astrophysics, Institute of High Energy Physics, Chinese Academy of Sciences, 19B Yuquan Road, Beijing 100049, China}
%
\author{Shi-Jie Zheng}
\affiliation{Key Laboratory of Particle Astrophysics, Institute of High Energy Physics, Chinese Academy of Sciences, 19B Yuquan Road, Beijing 100049, China}
%
\author{Xin-Qiao Li}
\affiliation{Key Laboratory of Particle Astrophysics, Institute of High Energy Physics, Chinese Academy of Sciences, 19B Yuquan Road, Beijing 100049, China}
%
\author{Xiang-Yang Wen}
\affiliation{Key Laboratory of Particle Astrophysics, Institute of High Energy Physics, Chinese Academy of Sciences, 19B Yuquan Road, Beijing 100049, China}
%
\author{Ke Gong}
\affiliation{Key Laboratory of Particle Astrophysics, Institute of High Energy Physics, Chinese Academy of Sciences, 19B Yuquan Road, Beijing 100049, China}
%
\author{Zheng-Hua An}
\affiliation{Key Laboratory of Particle Astrophysics, Institute of High Energy Physics, Chinese Academy of Sciences, 19B Yuquan Road, Beijing 100049, China}
%
\author{Da-Li Zhang}
\affiliation{Key Laboratory of Particle Astrophysics, Institute of High Energy Physics, Chinese Academy of Sciences, 19B Yuquan Road, Beijing 100049, China}
%
\author{Sheng Yang}
\affiliation{Key Laboratory of Particle Astrophysics, Institute of High Energy Physics, Chinese Academy of Sciences, 19B Yuquan Road, Beijing 100049, China}
%
\author{Xiao-Jing Liu}
\affiliation{Key Laboratory of Particle Astrophysics, Institute of High Energy Physics, Chinese Academy of Sciences, 19B Yuquan Road, Beijing 100049, China}
%
\author{Fan Zhang}
\affiliation{Key Laboratory of Particle Astrophysics, Institute of High Energy Physics, Chinese Academy of Sciences, 19B Yuquan Road, Beijing 100049, China}

\begin{abstract}
Magnetar is a neutron star with an ultrahigh magnetic field ($\sim 10^{14}-10^{15}$ G). The magnetar SGR J1935+2154 is not only one of the most active magnetars detected so far, but also the unique confirmed source of fast radio bursts (FRBs).
Gravitational wave high-energy Electromagnetic Counterpart All-sky Monitor (GECAM) is dedicated to monitor gamma-ray transients all over the sky, including magnetar short bursts.
Here we report the GECAM observations of the burst activity of SGR J1935+2154 from January 2021 to December 2022, which results in a unique and valuable data set for this important magnetar. With a targeted search of GECAM data, 159 bursts from SGR J1935+2154 are detected by GECAM-B while 97 bursts by GECAM-C, including the X-ray burst associated with a bright radio burst.
We find that both the burst duration and the waiting time between two successive bursts follow lognormal distributions. The period of burst activity is $134\pm20$ days, thus the burst activity could be generally divided into four active episodes over these two years. 
Interestingly, the hardness ratio of X-ray bursts tends to be softer during these two years, especially during the active episode with radio bursts detected.
\end{abstract}


\section{Introduction} \label{sec:intro}
Magnetars are highly magnetized isolated neutron stars \citep{ThompsonDuncan1992,vanKerkwijk1995,Kouveliotou1998,Banas1997,KaspiBeloborodov2017}, and are characterized by their slow rotation period (P$\sim2-12$ s), rapidly spinning down ($\dot{P} \sim 10^{-13}-10^{-11}$ $\mathrm{s \cdot s^{-1}}$), and relatively young age (typically about 1000 yr).
We currently know about 30 magnetars \citep{Olausen2014}\footnote{\url{http://www.physics.mcgill.ca/~pulsar/magnetar/main.html}}.
Magnetars are manifested themselves by persistent emission and emit bursts/flares simultaneously in the X-/gamma-ray band \citep{Esposito2021ASSL}.

Based on their luminosity and duration, magnetar bursts can be divided into three classes \citep{WoodsThompson2006}: Short-duration burst, which consists of single or multiple peaks, is the most typical magnetar burst with a burst duration $\sim$0.01 $\mathrm{s}$ and the luminosity around $10^{38} \sim 10^{40}$ $\mathrm{erg \cdot s^{-1}}$; Intermediate burst is a brighter magnetar burst with a duration longer than a short-duration burst ($> 1$ $\mathrm{s}$) and peak luminosity around $10^{41} \sim 10^{43}$ $\mathrm{erg \cdot s^{-1}}$; Giant flare, the rarest and the most powerful energetic burst is characterized by a significantly higher luminosity than a typical magnetar burst and a light curve with a hard initial spike and rapidly decaying tail modulated at the spin period of the magnetar.

Magnetar SGR J1935+2154 was first discovered and located in the Milky Way Galaxy by the Swift Burst Alert Telescope (BAT) in 2014 July \citep{Stamatikos2014}, while it is worth noting that \cite{Xie2022mnras} found several burst candidates in the gamma-ray burst monitor of Fermi gamma-ray Space Telescope \citep[Fermi/GBM,][]{Meegan2009} before this time. Follow-up observations carried out between 2014 July and 2015 March with Chandra and XMM-Newton allowed the measurement of its spin period and spin-down rate, $P \sim 3.24 \ \mathrm{s}$ and $\dot{P} \sim 1.43 \times 10^{-11} \ \mathrm{s \cdot s^{-1}}$, respectively. This indicates the estimated magnetic field at the pole is $B \sim 2.2 \times 10^{14} \ \mathrm{G}$ \citep{Israel2016}. 
It has experienced multiple outbursts from 2014 to 2022 \citep{Younes2017,Lin2020a,Lin2020b,Cai2022apjs_A,Cai2022apjs_b,Xie2022mnras,Rehan2023ApJ,Rehan2024ApJ}. 
April 2020 was recognized as a month of intense bursting activity for SGR J1935+2154, during which burst forests were observed \citep{Kaneko2021ApJ,Younes2020ApJ}. Moreover, the source entered a new outburst \citep{Borghese2022MNRAS}.
Among these bursts there is an X-ray burst \citep{Li2021Nat,Mereghetti2020,Ridnaia2021NatAs,Tavani2020ATel} associated with a fast radio burst, FRB 200428 \citep{Bochenek2020,CHIMEFRBCollaboration2020a}.
Additional coincident radio and X-ray bursts were observed during the active episode of 2022 \citep{LiWen2021,XiaoXiong2022mnras,Zhang2023nimpra,Wang2024ea,Wang2022ATel,Li2022ATel}.

In this work, we carry out a comprehensive targeted search on X-ray bursts of SGR J1935+2154 using GECAM observations data from January 2021 to December 2022.
As in Paper I of this series work for GECAM observations of SGR J1935+2154, we focus on the burst search, identification, and temporal analyses. In Paper II (in prep.), we will give a detailed time-integrated spectral analysis of these bursts.
The current paper is organized as follows: In Section \ref{sec:obs_bs}, we present the observations, the search process, and report the burst catalog. In Section \ref{sec:cata_ana}, we present the catalog analysis method and the results. Finally, a summary and discussions are given in Section \ref{sec:summa_discus}.

Please note that the error reported in this work is for 1 sigma, if not stated otherwise.

\section{Observation and Burst Search} \label{sec:obs_bs}
As of the writing of this paper, GECAM constellation is composed of four instruments, GECAM-A/B \citep[were launched in December 2020,][]{xiaoLiu2022mnras}, GECAM-C \citep[i.e. SATech-01/HEBS, launched in July 2022,][]{Zhang2023nimpra}, and GECAM-D \citep[i.e. DRO/GTM, launched in March 2024,][]{Wang2024ea}.
These four satellites are all-sky monitors with a large field of view that could monitor various gamma-ray transients except for the area blocked by Earth.
Both GECAM-A and GECAM-B feature a dome-shaped array of 25 gamma-ray detectors (GRDs) and 8 charged particle detectors (CPDs), while GECAM-C has 12 GRDs and 2 CPDs. All GECAM-B's GRDs and most GECAM-C's GRDs operate in two readout channels: high gain (HG, 6–300 keV) and low gain (LG, 300 keV–5 MeV), which are independent in terms of data processing, transmission, and dead time.
The event-by-event (EVT) data from GECAM's GRD detectors is used for further analysis. The GECAM achieves a high time resolution of 0.1 $\mathrm{\mu s}$, with a dead time of 4 $\mathrm{\mu s}$ for normal events and exceeding 69 $\mathrm{\mu s}$ for overflow events \citep{Liu2021arXiv}. We note that in this work only GECAM-B/C data are used because of power supply issues that GECAM-A has not been able to observe the source yet \citep{Li2021hzr}.

\subsection{Search and Identification}
Magnetar SGR J1935+2154 experienced multiple active X-ray burst episodes after the launch of GECAM-A/B, and an active X-ray burst episode after the launch of GECAM-C \citep{Rehan2023ApJ,Rehan2024ApJ,Wood2022GCN}. A search pipeline developed by \cite{Cai2024SCP} for gamma-ray bursts (GRBs) is used to do a blind search on GECAM daily observation data.
In addition to other GRBs, 82 X-ray bursts from SGR J1935+2154 were also found by GECAM-B \citep[2021-01 to 2022-01,][]{Xie2022mnras}\footnote{The 94 bursts shown in \cite{Xie2022mnras} are the preliminary blind searched results that are reported in the Gamma-ray Coordinates Network (GCN) by the burst advocate (BA). We identify 82 bursts from SGR J1935+2154 using the targeted search in this paper.}.

The search pipeline is implemented using a ground sensitive coherent search method that features a log-likelihood ratio (hereafter LR) comparing expected model counts to observed counts in each channel of each detector \citep{Blackburn2015ApJS,Cai2024SCP},
\begin{equation}
    \mathcal{L}=\sum_{i=1}^j[\ln\frac{\sigma_{n_\mathrm{i}}}{\sigma_{d_\mathrm{i}}}+\frac{\widetilde{d_\mathrm{i}^2}}{2\sigma_{n_\mathrm{i}}^2}-\frac{(\widetilde{d_\mathrm{i}}-r_\mathrm{i}s)^2}{2\sigma_{d_\mathrm{i}}^2}],
\end{equation}
where $i$ is the number of data sets in each channel and detector, $j$ is the total number of detectors and channels, $\widetilde{d_\mathrm{i}^2}$ is the background-subtracted counts and $\sigma_{d_\mathrm{i}}$ is the error of the expected counts (background+source), $n_\mathrm{i}$ is the estimated background and $\sigma_{n_\mathrm{i}}$ is the error of the background counts, $r_\mathrm{i}$ and $s$ represent the instrument response and the expected source amplitude (could be estimated by a spectral model template), respectively. Therefore, the LR could be calculated with the response factors over the all-sky for a blind search or a certain area for a targeted search. Then, the continuous time-bins with the higher LR (higher than the LR threshold) could be considered as one signal (trigger/burst). The maximum LR gives the most possible spatial location of the source signal. The pipeline will mark the first time-bin, of which LR is higher than the LR threshold, as a trigger time. In this paper, the LR threshold is set to 20 and the search timescale (time-bin) is set to 10 ms.

To unveil potential bursts that are relatively weak and unable to trigger the instrument with a normal threshold (typically $5\sigma$ as implemented in flight software), we conduct a targeted sub-threshold search for the area around the position of SGR J1935+2154 (R.A. and Decl. $\pm 20^\circ$) on GECAM daily EVT data\footnote{\url{https://gecamweb.ihep.ac.cn/dailydatadownload.jhtml}} from 2021-01-01 to 2022-12-31 using the pipeline developed by \cite{Cai2024SCP}.
Since X-ray bursts from magnetars tend to be softer than general GRBs, it is essential to employ appropriate spectral model templates instead of default Band templates for the search. Based on previous studies on magnetar burst spectral properties \citep{Lin2020a,Lin2020b,Rehan2023ApJ,Rehan2024ApJ}, we utilize the following spectral model templates: the soft Band function (default softer template for GRB searching), CPL (a power law with high-energy exponential cutoff), OTTB (optically-thin thermal bremsstrahlung), Blackbody, and Powerlaw. The parameter setting is listed in Table \ref{tab:template} which is similar to Table 1 in \cite{Xie2022mnras}. The details of each template are presented below.

The Band function:
\begin{equation}
A(E)\sim\begin{cases}E^{\alpha}\exp\left[-\frac{(2+\alpha)E}{E_\mathrm{p}}\right],&\text{if }E<\frac{(\alpha-\beta)E_\mathrm{p}}{2+\alpha},\\\\E^{\beta}\exp(\beta-\alpha)\left[\frac{(\alpha-\beta)E_\mathrm{p}}{(2+\alpha)}\right]^{\alpha-\beta},&\text{otherwise,}\end{cases}
\end{equation}
where $E_\mathrm{p}$, in units of keV, is the peak energy of the $\nu F_{\mathrm{\nu}}$ spectrum, $\alpha$/$\beta$ are the low-/high-energy photon indices.

The CPL function:
\begin{equation}
A(E) \sim E^{-\alpha}\exp \left[-\frac{(2-\alpha)E}{E_\mathrm{p}}\right],
\end{equation}
where $E_\mathrm{p}$, in units of keV, is the peak energy of the $\nu F_{\mathrm{\nu}}$ spectrum, $\alpha$ are the photon index.

The OTTB function:
\begin{equation}
A(E) \sim E^{-1} \exp\left(-\frac{E}{kT}\right),
\end{equation}
where $kT$ is the electron energy, in units of keV.

The Blackbody function:
\begin{equation}
    A(E) \sim \frac{E^2}{\exp(E/kT)-1},
\end{equation}
where $kT$ is the temperature, in units of keV.

The Powerlaw function:
\begin{equation}
    A(E) \sim E^{-\alpha},
\end{equation}
where $\alpha$ are the photon index.

Thousands of triggers (candidate bursts) have been found through the targeted search process. We perform a refined analysis for each candidate burst by carefully removing various false triggers or other burst events using the GRB data analysis tool, \textit{GECAMTools}\footnote{\url{https://github.com/zhangpeng-sci/GECAMTools-Public}}.
SGR J1935+2154 burst identified from the candidates should adhere to the following criteria \citep[same as][]{Xie2022mnras}:

\begin{enumerate}
\item Using the GECAM position history file (posatt) to retrieve the position of satellite and the pointing of detector. The candidates are excluded when SGR J1935+2154 is occulted by Earth;
\item The burst location is within the error circle (at 3 $\sigma$) of the source position \citep[SGR J1935+2154, R.A. = 293.73°, Decl. = 21.90°,][]{Israel2016};
\item The solar flares, the particle events measured by CPD observations, and the Terrestrial gamma-ray flash \citep[TGF,][]{Briggs2013} and Terrestrial electron beam \citep[TEB,][]{Xiong2012} featured by being harder ($\sim$ MeV) and shorter ($\sim$ 1 ms) than magnetar bursts are excluded;
\item The pipeline will calculate the significance\footnote{In this paper, the significance is defined as $S/\sqrt{B}$, where $S$ is the source (background-subtracted) counts and $B$ is the estimated background counts.} of each detector. At least two most significant detectors should have an angle relative to the source position of less than $60^\circ$ or maybe a false trigger (e.g., particle events);
\item Identified by other instruments (e.g., Fermi/GBM).
\end{enumerate}

\movetableright=-1cm
\begin{table}
\caption{Spectral templates used in the targeted search for X-ray bursts from SGR J1935+2154.}
\label{tab:template}
\begin{tabular}{lccc}
\hline
Spectrum Template & alpha & beta & Epeak or kT (keV) \\  \hline
SoftBand          & -1.9  & -3.7 & 70    \\
CPL               & 0.25  & ...  & 28    \\
OTTB              & ...   & ...  & 25    \\
Blackbody         & ...   & ...  & 8     \\
Powerlaw          & 2.4  & ...  & ...   \\
\hline
\end{tabular}
\end{table}

\begin{table}
\caption{The proportion of different templates in X-ray bursts that GECAM-B/C observes from SGR J1935+2154.}
\label{tab:temp_frac}
\begin{tabular}{cccc}
\hline
Instrument & Template & Number & Proportion \\
\hline
GECAM-B & SoftBand  & 27 & 16.98\% \\
        & CPL       & 39 & 24.52\% \\
        & OTTB      & 55 & 34.59\% \\
        & Blackbody & 29 & 18.23\% \\
        & Powerlaw  & 9 &  5.66\% \\
\hline
GECAM-C & SoftBand  & 7  &  7.21\% \\
        & CPL       & 35 & 36.08\% \\
        & OTTB      & 24 & 24.74\% \\
        & Blackbody & 19 & 19.58\% \\
        & Powerlaw  & 12 & 12.37\% \\
\hline
All bursts  & SoftBand  & 34 & 13.28\% \\
        & CPL       & 74 & 28.91\% \\
        & OTTB      & 79 & 30.86\% \\
        & Blackbody & 48 & 18.75\% \\
        & Powerlaw  & 21 & 8.20\% \\
\hline
\end{tabular}
\end{table}

\subsection{Search Results}
A total of 159 bursts for GECAM-B and 97 bursts for GECAM-C are identified and listed in Tables \ref{tab:burst_list_gb} and \ref{tab:burst_list_gc}.
Among them, there are 14 bursts that both GECAM-B and GECAM-C could observe, and a short X-ray burst (2022-10-14T19:21:39.130 UTC) observed by GECAM-B\&C is associated with a bright radio burst FRB 20221014 \citep{DongChime2022ATel,WangXiong2022ATel,Frederiks2022ATel,Maan2022ATel,Giri2023arXiv}\footnote{We mark this radio burst as FRB 20221014 in this paper since the luminosity is several orders higher than the pulsed emission from SGR J1935+2154 \citep{Zhu2023SciA} although still a few orders lower than that of general FRBs.}.
It is worth noting that about one-third of GECAM bursts (i.e. 60 bursts of GECAM-B and 36 bursts of GECAM-C) are invisible to Fermi/GBM\footnote{Using the Fermi/GBM position history file (poshist) and the GBM data tool \citep{GbmDataTools} to determine if SGR J1935+2154 is visible to Fermi/GBM or occulted by the Earth at these burst times.}.

\begin{figure*}
\epsscale{1.0}
\plottwo{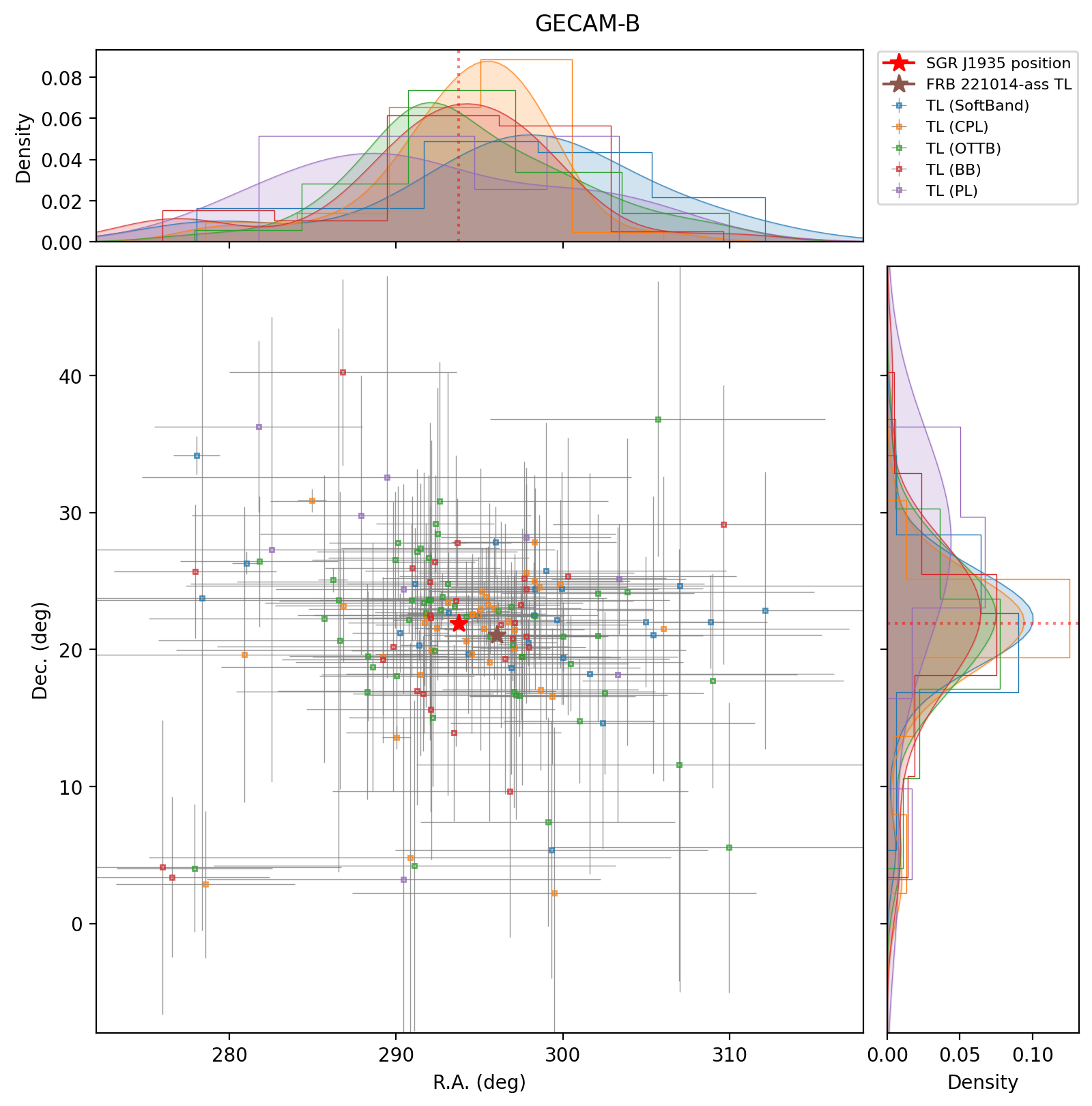}{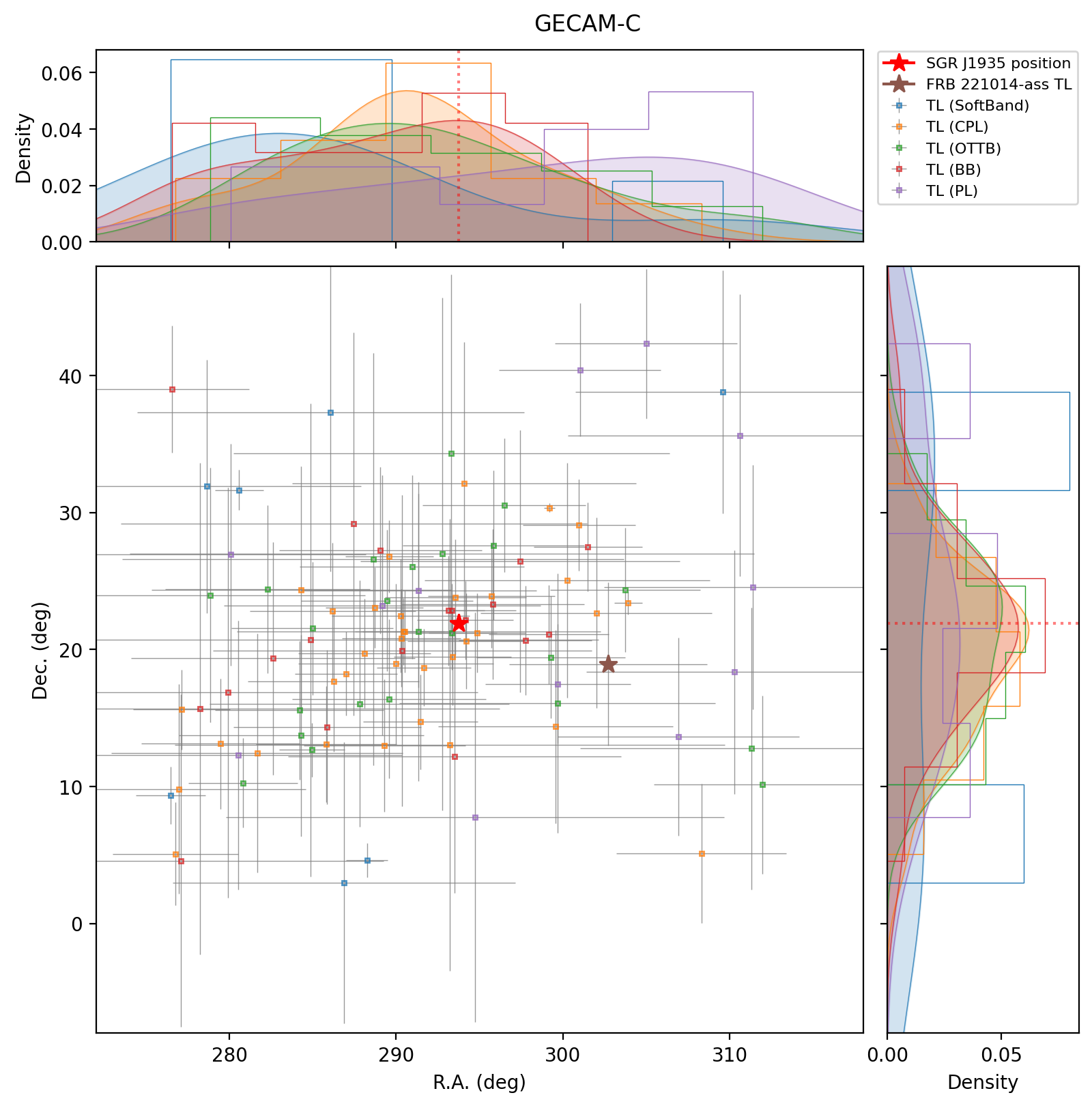}
\caption{The targeted search location (TL) of bursts. The red star marks the accurate position of SGR J1935+2154 (R.A. = 293.73°, Decl. = 21.90°).}
\label{fig:loc}
\end{figure*}

\begin{figure*}
\epsscale{1.0}
\plottwo{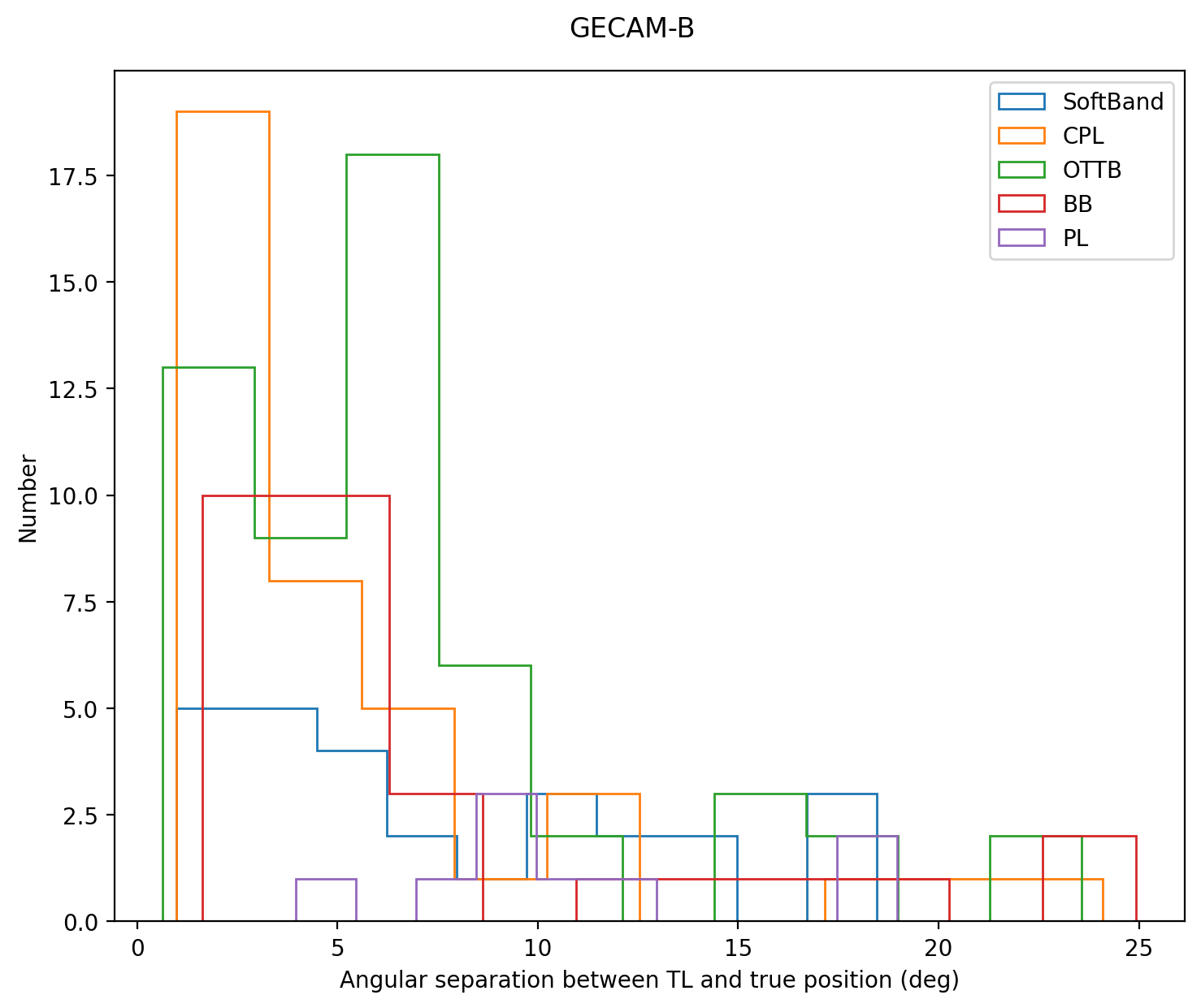}{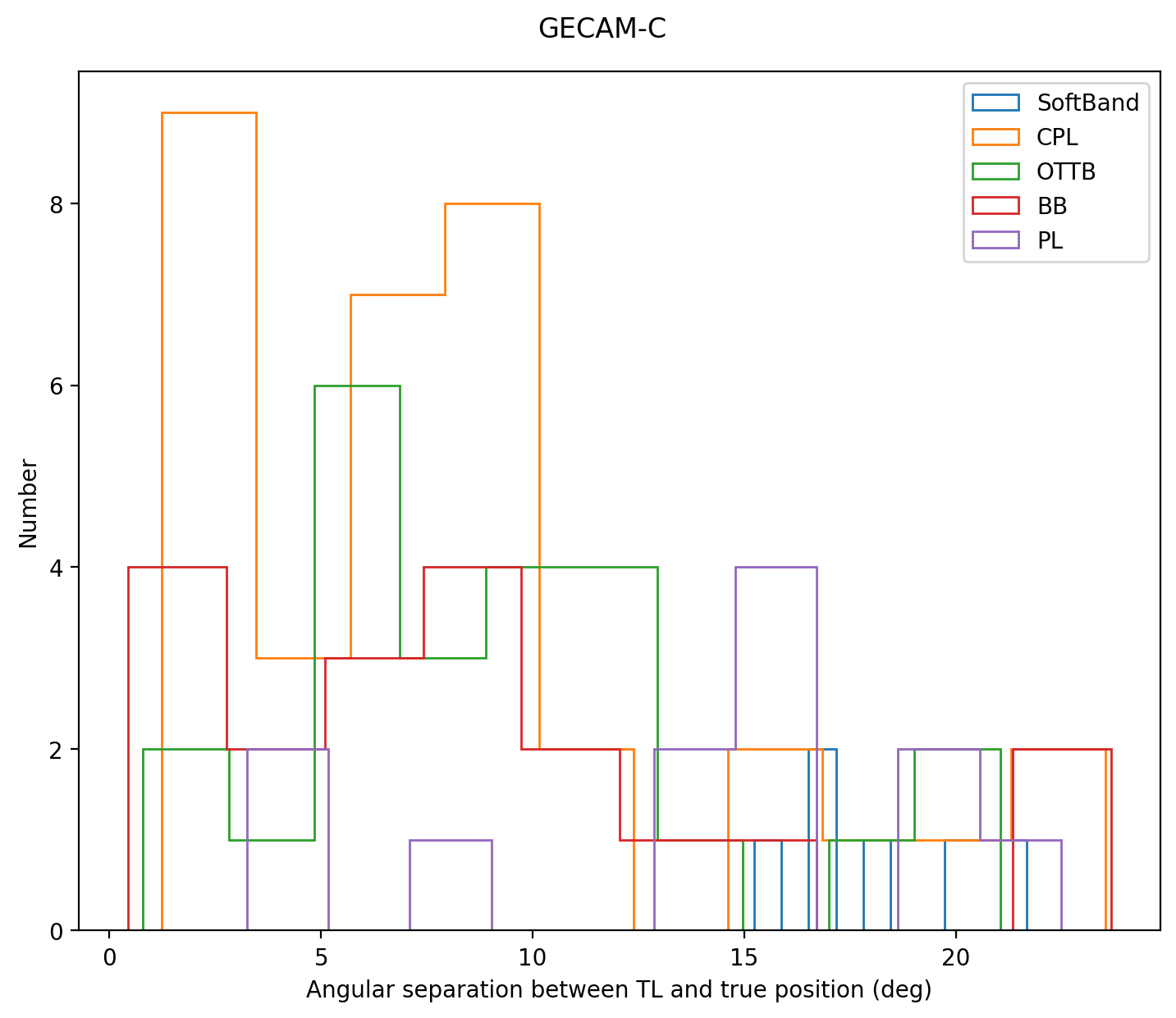}
\caption{The angle separation between the targeted search location (TL) of burst and true position of SGR J1935+2154.}
\label{fig:dist_ang_sep}
\end{figure*}

\begin{figure*}
\epsscale{1.0}
\plottwo{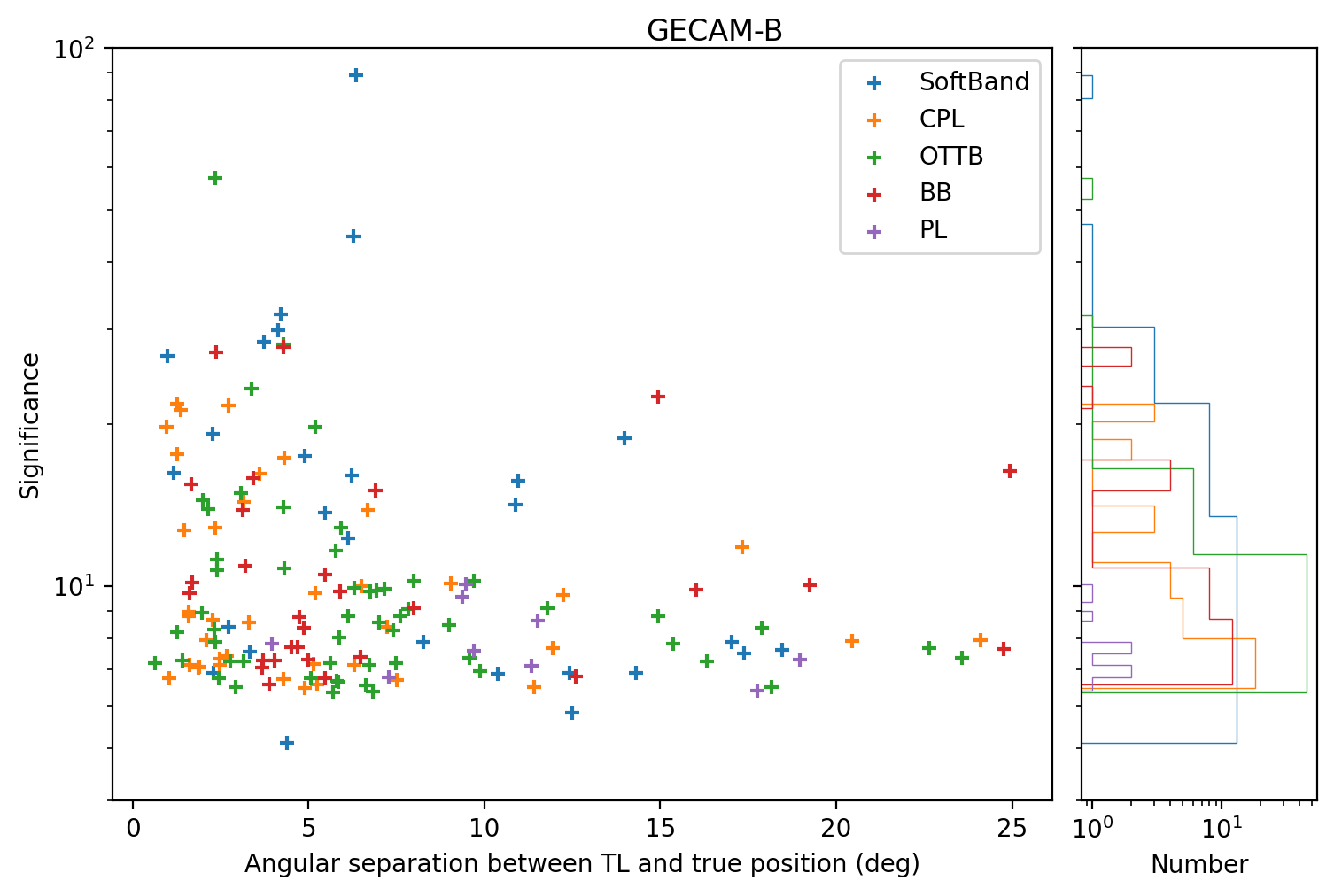}{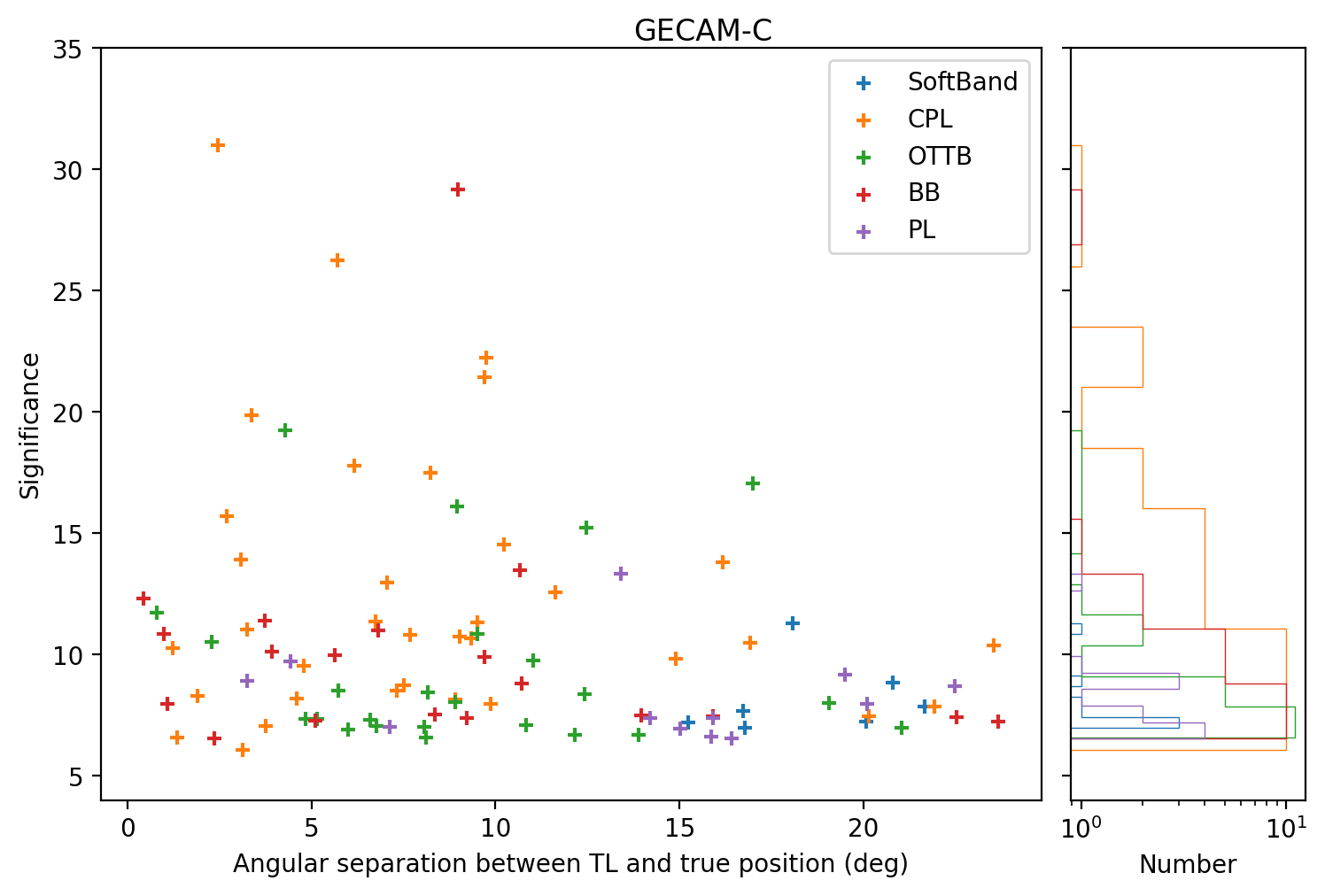}
\caption{The significance of burst v.s angle separation between the targeted search location (TL) and true position of SGR J1935+2154.}
\label{fig:ang_sep_sig}
\end{figure*}

The targeted search location (TL) of each burst is plotted in Fig \ref{fig:loc}. We summarize the proportion of different templates in Table \ref{tab:temp_frac} and Fig. \ref{fig:dist_ang_sep} to assess each template's merit for magnetar bursts to the targeted search. The CPL, OTTB, and BB account for a higher proportion of them than other templates.
Note that even a burst could be found by every template, the pipeline only outputs the best template \citep[e.g., the one with the maximum log-likelihood ratio, as referenced in][]{Cai2021mnras}.
As shown in Fig. \ref{fig:ang_sep_sig}, using a Powerlaw template to search for a magnetar burst prefers a weak burst with low significance. Most of the SoftBand templates are given by GECAM-B data after October 2022, showing a large difference compared to those from GECAM-C. This disparity may be attributed to the effective detectable energy range of GECAM-B changed in the HG channel ($>$40 keV after October 2022), potentially leading to bursts with more high energy photons being detected without softer photons, making them easier to find using the SoftBand compared to other templates.
Therefore, in general, it is recommended to use a template with a simple power-law function and exponential cutoff function, such as CPL, OTTB, or Blackbody, to coherently search for and locate magnetar bursts in hard X-ray observations.

\startlongtable
\begin{deluxetable*}{ccccccccccc}
\tabletypesize{\footnotesize}
\tablecaption{SGR J1935+2154 burst list observed by GECAM-B from January 2021 to December 2022.}\label{tab:burst_list_gb}
\tablehead{
\colhead{Burst ID} & \colhead{Trigger Time} & \colhead{Template} & \colhead{R.A.} & \colhead{Decl.} & \colhead{Err} & \colhead{Sig.} & \colhead{$T_\mathrm{bb}$\tablenotemark{a}} & \colhead{$T_\mathrm{st}$\tablenotemark{b}} & \colhead{Visible to} & \colhead{Visible to}\\
\colhead{} & \colhead{(UTC)} & \colhead{} & \colhead{(deg)} & \colhead{(deg)} & \colhead{(deg)} & \colhead{($\sigma$)} & \colhead{(ms)} & \colhead{(ms)} & \colhead{GBM} & \colhead{GECAM-C}
}
\startdata
1 & 2021-01-02T03:15:43.330 & OTTB & 277.91 & 4.04 & 4.66 & 7.35 & 4.73 & -2.58 & Yes & No \\
2 & 2021-01-06T05:11:16.390 & BB & 276.55 & 3.39 & 5.85 & 16.33 & 18.63 & -8.80 & Yes & No \\
3 & 2021-01-24T00:00:59.050 & CPL & 290.82 & 4.82 & 15.64 & 11.82 & 13.39 & -10.20 & Yes & No \\
4 & 2021-01-25T03:09:50.100 & PL & 290.44 & 3.23 & 11.80 & 7.30 & 130.96 & -84.19 & Yes & No \\
5 & 2021-01-26T18:59:44.750 & BB & 297.66 & 25.20 & 8.51 & 8.35 & 38.18 & -7.44 & Yes & No \\
6 & 2021-01-27T06:50:20.750 & CPL & 292.43 & 21.57 & 1.36 & 21.78 & 122.33 & -36.92 & No & No \\
7 & 2021-01-29T17:51:00.800 & PL & 303.28 & 18.16 & 10.76 & 7.59 & 269.71 & -170.50 & Yes & No \\
8 & 2021-01-30T08:39:53.840 & PL & 297.79 & 28.21 & 5.09 & 6.76 & 147.77 & -28.85 & Yes & No \\
9 & 2021-01-30T08:48:23.580 & OTTB & 292.49 & 28.45 & 10.66 & 6.53 & 6.72 & -3.53 & Yes & No \\
10 & 2021-01-30T10:35:35.120 & BB & 286.80 & 40.23 & 6.82 & 10.03 & 19.55 & 0.62 & Yes & No \\
11 & 2021-01-30T17:40:54.750 & OTTB & 292.67 & 22.92 & 1.35 & 7.27 & 121.16 & -0.74 & Yes & No \\
12 & 2021-01-30T21:01:22.865 & OTTB & 286.55 & 23.59 & 19.83 & 6.37 & 72.04 & -5.88 & Yes & No \\
13 & 2021-02-11T13:43:16.760 & BB & 275.98 & 4.12 & 10.73 & 7.64 & 28.50 & -24.53 & Yes & No \\
14 & 2021-02-16T22:20:39.600 & CPL & 290.02 & 13.60 & 0.85 & 10.11 & 301.80 & 17.46 & Yes & No \\
15 & 2021-07-07T00:33:31.640 & OTTB & 292.75 & 23.87 & 1.68 & 13.88 & 124.74 & 10.48 & Yes & No \\
16 & 2021-07-08T00:18:18.560 & SoftBand & 278.35 & 23.77 & 24.26 & 6.90 & 302.37 & -25.77 & Yes & No \\
17 & 2021-07-10T05:05:41.415 & CPL & 299.37 & 16.61 & 4.98 & 6.69 & 53.12 & -5.03 & No & No \\
18 & 2021-07-12T04:32:39.600 & OTTB & 289.92 & 26.53 & 4.98 & 11.62 & 22.23 & -32.91 & No & No \\
19 & 2021-07-12T22:12:58.100 & BB & 291.62 & 16.81 & 4.23 & 6.73 & 51.78 & -68.85 & No & No \\
20 & 2021-09-09T20:58:35.600 & BB & 309.61 & 29.12 & 10.20 & 9.86 & 28.62 & -39.05 & No & No \\
21 & 2021-09-09T21:07:12.150 & CPL & 295.39 & 23.87 & 3.57 & 7.32 & 55.13 & 10.39 & No & No \\
22 & 2021-09-10T01:04:33.500 & OTTB & 291.10 & 4.21 & 12.06 & 8.36 & 13.16 & 116.38 & Yes & No \\
23 & 2021-09-10T02:07:56.700 & OTTB & 288.57 & 18.71 & 9.07 & 6.66 & 71.03 & -48.93 & No & No \\
24 & 2021-09-10T02:08:28.800 & OTTB & 291.44 & 27.38 & 4.75 & 6.64 & 145.70 & -54.09 & No & No \\
25 & 2021-09-10T03:22:40.550 & CPL & 295.26 & 21.50 & 3.48 & 12.71 & 377.13 & -19.89 & No & No \\
26 & 2021-09-10T03:24:47.150 & OTTB & 297.37 & 16.63 & 2.97 & 9.91 & 48.37 & -9.69 & No & No \\
27 & 2021-09-10T03:42:45.750 & CPL & 280.88 & 19.63 & 10.79 & 9.63 & 52.04 & -55.59 & No & No \\
28 & 2021-09-10T05:05:03.350 & OTTB & 302.51 & 16.83 & 5.93 & 10.22 & 302.41 & -111.45 & No & No \\
29 & 2021-09-10T05:35:55.500 & OTTB & 302.09 & 24.12 & 5.78 & 10.21 & 21.73 & -17.50 & Yes & No \\
30 & 2021-09-11T05:32:38.620 & BB & 292.32 & 26.42 & 2.39 & 7.70 & 221.50 & -17.22 & Yes & No \\
31 & 2021-09-11T16:35:46.500 & BB & 296.83 & 9.68 & 10.67 & 6.80 & 42.40 & -69.29 & Yes & No \\
32 & 2021-09-11T16:39:21.000 & OTTB & 299.09 & 7.43 & 7.63 & 7.82 & 53.30 & -86.96 & Yes & No \\
33 & 2021-09-11T16:50:03.850 & BB & 293.68 & 27.82 & 3.20 & 9.78 & 39.63 & -25.12 & Yes & No \\
34 & 2021-09-11T17:01:10.800 & CPL & 284.93 & 30.88 & 0.86 & 7.67 & 1349.53 & -1050.55 & Yes & No \\
35 & 2021-09-11T17:04:29.800 & BB & 289.81 & 20.21 & 10.56 & 7.28 & 16.48 & -49.93 & Yes & No \\
36 & 2021-09-11T17:10:48.750 & OTTB & 300.98 & 14.80 & 4.52 & 6.95 & 6.83 & 194.27 & Yes & No \\
37 & 2021-09-11T18:02:13.500 & OTTB & 309.94 & 5.57 & 10.59 & 7.66 & 38.60 & -37.01 & Yes & No \\
38 & 2021-09-11T18:04:46.350 & CPL & 297.80 & 25.61 & 6.46 & 6.57 & 61.00 & -28.29 & Yes & No \\
39 & 2021-09-11T18:54:36.050 & BB & 296.30 & 21.81 & 7.54 & 27.16 & 37.25 & -20.13 & Yes & No \\
40 & 2021-09-11T19:43:28.000 & OTTB & 297.52 & 19.48 & 4.57 & 13.98 & 243.94 & -83.05 & Yes & No \\
41 & 2021-09-11T19:46:50.050 & OTTB & 297.55 & 19.50 & 5.33 & 10.77 & 77.06 & -84.11 & Yes & No \\
42 & 2021-09-11T20:13:40.550 & BB & 292.05 & 22.33 & 3.39 & 9.69 & 152.69 & -63.17 & Yes & No \\
43 & 2021-09-11T20:22:59.050 & CPL & 292.04 & 22.33 & 5.15 & 7.13 & 180.72 & -21.73 & Yes & No \\
44 & 2021-09-11T20:33:14.550 & PL & 281.74 & 36.29 & 6.24 & 6.39 & 87.57 & -105.27 & No & No \\
45 & 2021-09-11T21:07:28.350 & CPL & 292.10 & 19.98 & 15.28 & 7.12 & 48.17 & -183.59 & Yes & No \\
46 & 2021-09-11T22:51:41.600 & CPL & 297.54 & 19.47 & 9.35 & 17.29 & 116.25 & -29.25 & Yes & No \\
47 & 2021-09-12T00:34:37.450 & CPL & 295.13 & 24.27 & 3.19 & 7.41 & 347.42 & -41.07 & Yes & No \\
48 & 2021-09-12T00:45:49.400 & BB & 297.81 & 24.43 & 4.28 & 7.68 & 32.26 & -14.45 & Yes & No \\
49 & 2021-09-12T05:14:07.950 & CPL & 298.65 & 17.10 & 5.87 & 13.87 & 110.85 & -124.44 & Yes & No \\
50 & 2021-09-12T05:44:17.050 & OTTB & 306.96 & 11.58 & 15.77 & 7.23 & 80.14 & -37.80 & Yes & No \\
51 & 2021-09-12T16:26:08.150 & OTTB & 308.99 & 17.73 & 7.83 & 8.80 & 66.08 & -76.48 & Yes & No \\
52 & 2021-09-12T22:16:36.200 & PL & 287.91 & 29.81 & 10.18 & 10.08 & 8.48 & -13.39 & No & No \\
53 & 2021-09-13T00:27:25.200 & OTTB & 297.19 & 16.71 & 9.10 & 8.78 & 252.68 & -31.43 & Yes & No \\
54 & 2021-09-13T19:51:33.350 & CPL & 294.58 & 22.60 & 4.40 & 6.75 & 161.79 & -183.78 & Yes & No \\
55 & 2021-09-14T11:10:36.250 & CPL & 294.53 & 22.56 & 3.91 & 19.75 & 94.94 & -41.84 & Yes & No \\
56 & 2021-09-14T14:15:42.900 & BB & 293.45 & 13.94 & 6.45 & 9.10 & 13.05 & -18.35 & Yes & No \\
57 & 2021-09-14T23:21:58.500 & OTTB & 293.08 & 24.79 & 15.42 & 6.48 & 131.71 & -100.90 & No & No \\
58 & 2021-09-14T23:26:34.050 & CPL & 298.30 & 27.84 & 4.91 & 8.41 & 30.67 & -25.93 & No & No \\
59 & 2021-09-15T02:35:47.350 & CPL & 299.49 & 2.24 & 12.11 & 7.90 & 46.36 & -56.07 & No & No \\
60 & 2021-09-15T02:39:25.700 & OTTB & 302.11 & 21.02 & 6.28 & 9.06 & 66.16 & -56.21 & No & No \\
61 & 2021-09-15T15:32:56.050 & CPL & 306.03 & 21.53 & 11.12 & 6.49 & 109.28 & -97.54 & Yes & No \\
62 & 2021-09-17T12:52:37.800 & CPL & 295.59 & 19.10 & 11.61 & 8.54 & 32.38 & -54.13 & Yes & No \\
63 & 2021-09-17T13:58:25.100 & SoftBand & 307.02 & 24.67 & 29.66 & 5.82 & 173.25 & -191.07 & Yes & No \\
64 & 2021-09-18T22:58:52.150 & OTTB & 305.69 & 36.83 & 10.05 & 6.48 & 86.02 & -89.76 & Yes & No \\
65 & 2021-09-22T02:39:10.200 & CPL & 295.06 & 22.95 & 10.30 & 8.96 & 114.19 & -36.91 & No & No \\
66 & 2021-09-22T20:12:16.500 & OTTB & 286.20 & 25.10 & 0.87 & 8.78 & 170.69 & -31.95 & No & No \\
67 & 2021-09-29T23:41:12.245 & BB & 292.03 & 24.98 & 11.59 & 15.86 & 18.85 & -5.05 & No & No \\
68 & 2021-09-30T01:31:06.165 & BB & 292.04 & 22.52 & 7.96 & 10.14 & 50.53 & -13.05 & Yes & No \\
69 & 2021-10-01T00:04:04.340 & BB & 300.29 & 25.34 & 10.11 & 15.07 & 29.50 & -6.45 & Yes & No \\
70 & 2021-10-07T11:57:07.700 & CPL & 295.51 & 23.25 & 4.32 & 7.93 & 310.13 & -129.76 & No & No \\
71 & 2021-11-01T23:13:41.950 & OTTB & 291.66 & 23.42 & 9.52 & 6.74 & 26.91 & -41.10 & No & No \\
72 & 2022-01-04T04:32:11.200 & BB & 291.23 & 16.98 & 8.33 & 10.50 & 35.71 & -55.83 & Yes & No \\
73 & 2022-01-05T06:01:31.450 & BB & 297.80 & 20.99 & 4.91 & 6.56 & 75.39 & -96.74 & Yes & No \\
74 & 2022-01-05T07:06:40.800 & CPL & 294.95 & 22.51 & 4.25 & 17.59 & 126.75 & -52.60 & Yes & No \\
75 & 2022-01-06T02:36:14.100 & OTTB & 298.33 & 22.47 & 1.66 & 28.11 & 51.67 & -54.52 & Yes & No \\
76 & 2022-01-08T14:41:46.900 & SoftBand & 296.90 & 18.66 & 7.73 & 5.10 & 76.18 & -31.01 & Yes & No \\
77 & 2022-01-09T07:39:10.700 & CPL & 298.56 & 24.55 & 5.29 & 7.16 & 35.87 & -18.96 & Yes & No \\
78 & 2022-01-10T02:57:16.825 & BB & 297.95 & 20.19 & 3.40 & 27.78 & 84.87 & -5.71 & Yes & No \\
79 & 2022-01-10T06:52:40.500 & CPL & 289.20 & 19.47 & 4.23 & 6.46 & 63.07 & -34.89 & No & No \\
80 & 2022-01-11T08:58:35.450 & BB & 296.97 & 20.80 & 3.07 & 10.91 & 171.34 & -110.70 & Yes & No \\
81 & 2022-01-12T01:03:46.900 & OTTB & 288.32 & 19.51 & 4.73 & 7.18 & 116.26 & -111.12 & Yes & No \\
82 & 2022-01-12T02:19:22.200 & CPL & 291.72 & 21.95 & 4.59 & 7.05 & 392.65 & -142.87 & Yes & No \\
83 & 2022-01-12T05:42:51.650 & CPL & 295.89 & 23.01 & 5.18 & 8.65 & 198.73 & -173.05 & Yes & No \\
84 & 2022-01-12T08:39:25.450 & OTTB & 291.94 & 23.59 & 0.44 & 7.86 & 979.51 & -183.09 & Yes & No \\
85 & 2022-01-12T17:57:08.500 & OTTB & 290.11 & 27.80 & 4.14 & 7.14 & 147.71 & -143.36 & Yes & No \\
86 & 2022-01-13T19:36:08.600 & CPL & 291.78 & 22.50 & 6.73 & 7.08 & 48.18 & -86.71 & Yes & No \\
87 & 2022-01-13T20:06:58.760 & OTTB & 292.59 & 30.85 & 10.13 & 8.45 & 37.51 & -11.38 & Yes & No \\
88 & 2022-01-13T20:14:58.600 & OTTB & 300.47 & 19.00 & 3.47 & 9.81 & 459.77 & -453.04 & Yes & No \\
89 & 2022-01-13T21:41:17.900 & CPL & 298.29 & 25.00 & 5.74 & 9.69 & 46.64 & -52.24 & Yes & No \\
90 & 2022-01-14T19:42:08.050 & CPL & 294.18 & 20.61 & 2.09 & 21.27 & 234.33 & 805.01 & Yes & No \\
91 & 2022-01-14T19:45:08.100 & CPL & 291.46 & 18.19 & 5.96 & 6.71 & 27.85 & -37.33 & Yes & No \\
92 & 2022-01-14T19:56:52.700 & OTTB & 296.11 & 22.79 & 0.69 & 57.43 & 382.90 & -265.83 & Yes & No \\
93 & 2022-01-14T20:06:07.400 & OTTB & 291.24 & 27.15 & 6.01 & 6.34 & 36.48 & -62.88 & No & No \\
94 & 2022-01-14T20:07:03.050 & OTTB & 291.78 & 22.66 & 1.10 & 8.92 & 315.07 & -6.13 & No & No \\
95 & 2022-01-14T20:12:45.300 & BB & 292.07 & 15.64 & 7.45 & 7.38 & 511.42 & -525.26 & No & No \\
96 & 2022-01-14T20:15:54.400 & OTTB & 290.77 & 22.18 & 0.96 & 7.23 & 279.46 & -131.00 & No & No \\
97 & 2022-01-14T20:21:05.150 & OTTB & 293.51 & 23.17 & 0.15 & 8.20 & 1552.57 & -95.36 & No & No \\
98 & 2022-01-14T20:22:53.955 & BB & 297.11 & 21.96 & 5.04 & 13.85 & 360.17 & -3.72 & No & No \\
99 & 2022-01-14T20:23:35.400 & BB & 297.49 & 23.27 & 1.41 & 7.26 & 197.65 & -153.85 & No & No \\
100 & 2022-01-14T20:26:50.300 & BB & 289.22 & 19.27 & 6.06 & 7.29 & 17.96 & -29.98 & No & No \\
101 & 2022-01-14T20:29:07.250 & OTTB & 296.99 & 20.38 & 0.86 & 23.31 & 629.51 & -127.86 & No & No \\
102 & 2022-01-14T20:31:49.900 & OTTB & 292.19 & 15.06 & 5.05 & 8.57 & 197.92 & -148.02 & No & No \\
103 & 2022-01-15T09:26:39.900 & SoftBand & 294.31 & 19.69 & 4.34 & 19.15 & 38.68 & -36.41 & Yes & No \\
104 & 2022-01-15T13:52:26.050 & CPL & 294.53 & 19.68 & 5.77 & 12.82 & 72.31 & -42.00 & No & No \\
105 & 2022-01-15T16:31:14.900 & CPL & 299.83 & 24.81 & 6.13 & 7.14 & 270.39 & -263.97 & Yes & No \\
106 & 2022-01-15T17:21:59.300 & OTTB & 295.64 & 20.99 & 1.28 & 14.43 & 246.44 & -28.19 & Yes & No \\
107 & 2022-01-16T10:48:37.650 & CPL & 296.68 & 22.09 & 3.38 & 21.66 & 281.15 & -51.90 & Yes & No \\
108 & 2022-01-17T01:27:12.720 & OTTB & 297.06 & 16.94 & 8.47 & 8.03 & 88.85 & -35.25 & Yes & No \\
109 & 2022-01-17T01:39:37.300 & OTTB & 303.85 & 24.20 & 11.21 & 7.34 & 136.90 & -92.61 & Yes & No \\
110 & 2022-01-20T18:52:48.950 & OTTB & 291.96 & 26.72 & 6.02 & 6.75 & 31.61 & -41.89 & No & No \\
111 & 2022-01-23T20:06:38.750 & CPL & 297.12 & 20.14 & 2.10 & 16.17 & 357.03 & -8.50 & No & No \\
112 & 2022-01-24T02:10:55.050 & CPL & 293.10 & 23.40 & 2.88 & 8.79 & 251.47 & -47.28 & No & No \\
113 & 2022-02-02T14:52:28.170 & BB & 296.53 & 19.32 & 9.85 & 7.04 & 30.11 & -8.93 & No & No \\
114 & 2022-02-06T05:15:45.635 & CPL & 286.83 & 23.23 & 4.08 & 10.00 & 43.00 & -7.87 & Yes & No \\
115 & 2022-05-21T23:32:58.800 & BB & 290.97 & 25.93 & 5.26 & 8.76 & 39.62 & -28.50 & Yes & No \\
116 & 2022-05-23T23:29:22.850 & OTTB & 285.67 & 22.26 & 10.49 & 7.19 & 90.88 & -22.11 & No & No \\
117 & 2022-05-24T17:10:18.550 & CPL & 297.09 & 21.45 & 4.20 & 14.34 & 30.38 & -36.51 & No & No \\
118 & 2022-10-11T14:11:36.850 & SoftBand & 308.83 & 22.03 & 3.41 & 18.79 & 35.22 & -49.06 & Yes & No \\
119 & 2022-10-12T12:47:04.400 & SoftBand & 290.23 & 21.22 & 2.88 & 7.55 & 113.52 & -15.72 & Yes & No \\
120 & 2022-10-12T14:22:47.000 & SoftBand & 302.37 & 14.62 & 9.13 & 15.70 & 22.80 & -24.19 & Yes & Yes \\
121 & 2022-10-12T14:40:50.450 & SoftBand & 299.91 & 24.47 & 8.50 & 16.04 & 41.38 & -22.76 & No & Yes \\
122 & 2022-10-12T14:43:19.950 & SoftBand & 295.94 & 27.86 & 2.59 & 44.72 & 184.74 & -60.42 & No & Yes \\
123 & 2022-10-12T14:44:45.600 & SoftBand & 294.85 & 22.46 & 0.48 & 16.26 & 848.21 & -53.68 & No & No \\
124 & 2022-10-12T14:54:37.950 & SoftBand & 301.62 & 18.24 & 14.63 & 7.86 & 9.94 & -51.20 & No & No \\
125 & 2022-10-12T15:42:31.250 & SoftBand & 278.03 & 34.16 & 1.39 & 7.60 & 395.20 & -20.92 & Yes & No \\
126 & 2022-10-12T15:45:10.150 & SoftBand & 281.00 & 26.32 & 0.83 & 6.88 & 393.21 & -196.87 & Yes & Yes \\
127 & 2022-10-12T22:34:56.900 & SoftBand & 299.32 & 5.40 & 9.37 & 7.50 & 29.84 & -43.19 & No & Yes \\
128 & 2022-10-13T00:00:53.450 & SoftBand & 291.38 & 20.32 & 1.10 & 8.41 & 348.81 & -92.94 & Yes & Yes \\
129 & 2022-10-14T06:47:24.100 & SoftBand & 299.64 & 22.19 & 5.12 & 13.70 & 242.99 & -150.79 & No & No \\
130 & 2022-10-14T07:12:28.800 & SoftBand & 300.00 & 19.45 & 3.44 & 89.16 & 214.46 & -74.52 & Yes & No \\
131 & 2022-10-14T07:24:42.150 & SoftBand & 298.97 & 25.74 & 10.83 & 12.28 & 76.77 & -24.94 & Yes & Yes \\
132 & 2022-10-14T11:27:32.750 & SoftBand & 298.24 & 22.50 & 3.87 & 31.96 & 25.27 & -28.72 & No & Yes \\
133 & 2022-10-14T13:21:36.400 & SoftBand & 291.14 & 24.79 & 4.09 & 28.45 & 164.56 & -62.66 & Yes & Yes \\
134 & 2022-10-14T17:04:16.650 & CPL & 278.57 & 2.86 & 5.35 & 7.92 & 33.32 & -36.02 & Yes & Yes \\
135 & 2022-10-14T19:21:39.100 & SoftBand & 296.04 & 21.09 & 4.38 & 6.88 & 185.32 & -56.10 & No & Yes \\
136 & 2022-10-16T21:14:58.450 & SoftBand & 305.40 & 21.07 & 10.12 & 14.14 & 27.16 & -55.42 & Yes & No \\
137 & 2022-10-16T21:20:52.250 & SoftBand & 297.92 & 20.53 & 1.70 & 29.83 & 103.84 & -17.82 & Yes & Yes \\
138 & 2022-10-17T15:40:10.400 & BB & 277.95 & 25.72 & 4.88 & 22.52 & 146.21 & -21.22 & No & No \\
139 & 2022-10-17T16:24:53.900 & BB & 293.60 & 23.57 & 10.62 & 15.46 & 49.42 & -36.06 & Yes & No \\
140 & 2022-10-22T01:41:18.050 & SoftBand & 298.31 & 24.43 & 7.37 & 17.47 & 19.45 & -44.49 & No & Yes \\
141 & 2022-10-22T04:44:52.700 & SoftBand & 293.12 & 22.71 & 4.41 & 26.81 & 38.10 & -24.84 & No & Yes \\
142 & 2022-11-08T19:48:49.900 & PL & 282.55 & 27.31 & 16.97 & 8.63 & 23.47 & -38.93 & Yes & No \\
143 & 2022-11-09T14:33:10.200 & OTTB & 291.93 & 23.64 & 3.99 & 10.70 & 376.85 & -97.67 & Yes & Yes \\
144 & 2022-11-09T16:06:08.620 & OTTB & 296.91 & 23.11 & 1.48 & 7.24 & 2505.18 & -44.20 & Yes & Yes \\
145 & 2022-11-09T17:46:18.375 & PL & 290.42 & 24.43 & 7.66 & 7.80 & 145.80 & -25.70 & Yes & Yes \\
146 & 2022-11-09T17:47:59.220 & OTTB & 294.18 & 22.41 & 4.03 & 7.18 & 320.31 & -23.75 & Yes & Yes \\
147 & 2022-11-09T17:56:52.715 & OTTB & 288.28 & 16.94 & 7.89 & 9.88 & 36.62 & -22.52 & Yes & Yes \\
148 & 2022-11-09T18:25:56.380 & OTTB & 281.78 & 26.43 & 4.74 & 9.08 & 93.27 & -12.64 & No & Yes \\
149 & 2022-11-10T06:39:03.075 & SoftBand & 312.13 & 22.87 & 10.13 & 7.87 & 129.89 & -78.53 & Yes & No \\
150 & 2022-11-10T09:54:23.100 & OTTB & 290.04 & 18.06 & 4.66 & 19.78 & 92.53 & -20.96 & No & Yes \\
151 & 2022-11-10T15:15:13.355 & OTTB & 286.61 & 20.67 & 10.87 & 9.75 & 23.17 & -9.98 & Yes & No \\
152 & 2022-11-10T18:10:56.200 & PL & 303.35 & 25.14 & 4.04 & 9.54 & 25.16 & -23.64 & No & No \\
153 & 2022-11-11T13:11:49.125 & OTTB & 292.34 & 29.21 & 3.54 & 8.27 & 10.56 & -2.46 & No & Yes \\
154 & 2022-11-12T00:21:33.900 & OTTB & 292.05 & 23.64 & 3.53 & 8.31 & 146.83 & -60.74 & No & Yes \\
155 & 2022-11-12T03:32:17.705 & OTTB & 292.32 & 19.92 & 5.00 & 11.18 & 11.13 & -5.18 & No & Yes \\
156 & 2022-11-15T17:14:34.860 & OTTB & 290.93 & 23.59 & 5.55 & 14.86 & 164.34 & -6.85 & No & Yes \\
157 & 2022-11-19T21:53:33.050 & PL & 289.43 & 32.60 & 14.68 & 7.10 & 35.58 & -48.63 & Yes & Yes \\
158 & 2022-11-20T11:25:04.850 & SoftBand & 304.95 & 22.04 & 4.77 & 6.86 & 283.82 & -85.17 & Yes & Yes \\
159 & 2022-12-13T06:57:10.650 & OTTB & 300.02 & 20.96 & 2.51 & 12.81 & 133.42 & -68.79 & Yes & Yes \\
\enddata
\tablenotetext{a}{Burst duration derived by the Bayesian blocks algorithm.}
\tablenotetext{b}{Burst start time relative to the trigger time using the Bayesian blocks algorithm.}
\end{deluxetable*}

\startlongtable
\begin{deluxetable*}{ccccccccccc}
\tabletypesize{\footnotesize}
\tablecaption{SGR J1935+2154 burst list observed by GECAM-C from October to December 2022.}\label{tab:burst_list_gc}
\tablehead{
\colhead{Burst ID} & \colhead{Trigger Time} & \colhead{Template} & \colhead{R.A.} & \colhead{Decl.} & \colhead{Err} & \colhead{Sig.} & \colhead{$T_\mathrm{bb}$\tablenotemark{a}} & \colhead{$T_\mathrm{st}$\tablenotemark{b}} & \colhead{Visible to} & \colhead{Visible to}\\
\colhead{} & \colhead{(UTC)} & \colhead{} & \colhead{(deg)} & \colhead{(deg)} & \colhead{(deg)} & \colhead{($\sigma$)} & \colhead{(ms)} & \colhead{(ms)} & \colhead{GBM} & \colhead{GECAM-B}
}
\startdata
1 & 2022-10-09T10:32:49.680 & OTTB & 311.94 & 10.14 & 6.50 & 6.99 & 6.73 & -2.26 & Yes & Yes \\
2 & 2022-10-11T10:41:16.655 & CPL & 276.75 & 5.08 & 3.76 & 10.39 & 51.70 & -6.90 & Yes & No \\
3 & 2022-10-12T12:47:04.425 & OTTB & 296.47 & 30.54 & 4.89 & 16.11 & 496.77 & -299.79 & Yes & Yes \\
4 & 2022-10-12T15:06:51.200 & SoftBand & 309.59 & 38.81 & 8.85 & 7.86 & 39.92 & -82.27 & No & Yes \\
5 & 2022-10-12T15:07:36.710 & BB & 295.79 & 23.31 & 5.47 & 6.55 & 651.24 & -8.37 & No & Yes \\
6 & 2022-10-12T15:08:48.575 & OTTB & 299.27 & 19.45 & 4.48 & 8.51 & 779.10 & -8.11 & No & Yes \\
7 & 2022-10-12T15:11:19.095 & BB & 297.75 & 20.67 & 3.99 & 10.13 & 297.69 & -163.07 & No & Yes \\
8 & 2022-10-12T15:14:04.100 & BB & 294.14 & 22.16 & 2.87 & 12.30 & 724.92 & -10.46 & Yes & No \\
9 & 2022-10-12T15:14:34.525 & CPL & 294.18 & 20.63 & 3.49 & 6.59 & 320.39 & -64.49 & Yes & No \\
10 & 2022-10-12T15:14:51.325 & BB & 297.43 & 26.44 & 9.57 & 9.96 & 44.96 & -41.58 & Yes & No \\
11 & 2022-10-12T15:15:58.465 & PL & 289.17 & 23.23 & 9.50 & 9.73 & 37.52 & -6.34 & Yes & No \\
12 & 2022-10-12T15:16:22.975 & CPL & 289.97 & 18.96 & 5.85 & 8.17 & 1219.09 & -36.12 & Yes & No \\
13 & 2022-10-12T15:16:45.925 & OTTB & 299.67 & 16.11 & 9.47 & 7.03 & 7.25 & -7.07 & Yes & No \\
14 & 2022-10-12T15:18:51.805 & CPL & 286.25 & 17.67 & 5.14 & 17.48 & 98.53 & -21.56 & Yes & No \\
15 & 2022-10-12T15:20:30.930 & BB & 293.34 & 22.84 & 1.98 & 10.83 & 2661.75 & -25.13 & Yes & No \\
16 & 2022-10-12T15:21:39.825 & CPL & 299.55 & 14.37 & 7.04 & 10.68 & 11.79 & -6.85 & Yes & No \\
17 & 2022-10-12T15:22:25.140 & CPL & 290.27 & 22.46 & 4.15 & 11.02 & 1117.16 & -8.40 & Yes & No \\
18 & 2022-10-12T15:49:13.545 & PL & 304.99 & 42.33 & 5.47 & 8.69 & 439.68 & -208.56 & Yes & Yes \\
19 & 2022-10-12T18:33:43.480 & CPL & 303.91 & 23.43 & 0.84 & 11.33 & 92.65 & -9.66 & Yes & No \\
20 & 2022-10-13T02:13:21.465 & OTTB & 280.81 & 10.28 & 3.26 & 17.04 & 148.57 & -37.56 & No & Yes \\
21 & 2022-10-13T13:59:10.900 & BB & 290.36 & 19.93 & 11.36 & 11.40 & 6.45 & -9.10 & Yes & No \\
22 & 2022-10-13T15:30:31.525 & CPL & 293.56 & 23.82 & 4.23 & 8.28 & 830.16 & -38.84 & Yes & No \\
23 & 2022-10-13T22:41:29.435 & CPL & 299.20 & 30.35 & 0.32 & 22.25 & 501.80 & -82.95 & No & Yes \\
24 & 2022-10-13T22:55:29.320 & BB & 293.50 & 12.22 & 9.99 & 9.89 & 27.75 & -11.30 & Yes & Yes \\
25 & 2022-10-14T02:27:47.140 & BB & 293.15 & 22.85 & 4.02 & 7.97 & 54.90 & -2.92 & Yes & Yes \\
26 & 2022-10-14T05:42:15.425 & OTTB & 293.30 & 34.32 & 13.08 & 8.35 & 111.34 & -26.66 & Yes & Yes \\
27 & 2022-10-14T05:47:52.600 & OTTB & 282.28 & 24.40 & 6.14 & 7.10 & 317.33 & -65.42 & Yes & Yes \\
28 & 2022-10-14T06:47:24.045 & OTTB & 287.80 & 16.05 & 8.99 & 6.58 & 265.49 & -102.01 & No & Yes \\
29 & 2022-10-14T07:12:28.740 & CPL & 291.66 & 18.70 & 2.81 & 7.06 & 215.53 & -13.67 & Yes & Yes \\
30 & 2022-10-14T07:17:36.155 & OTTB & 289.57 & 16.41 & 5.80 & 7.06 & 306.75 & -18.63 & Yes & Yes \\
31 & 2022-10-14T09:56:04.850 & CPL & 308.30 & 5.13 & 5.09 & 7.84 & 149.19 & -96.41 & No & Yes \\
32 & 2022-10-14T10:28:39.825 & OTTB & 290.95 & 26.03 & 6.72 & 7.36 & 139.10 & -33.50 & Yes & Yes \\
33 & 2022-10-14T11:21:05.325 & CPL & 284.30 & 24.38 & 8.98 & 10.75 & 136.81 & -44.76 & No & Yes \\
34 & 2022-10-14T11:27:32.730 & CPL & 293.39 & 19.48 & 3.52 & 30.99 & 44.95 & -13.14 & No & Yes \\
35 & 2022-10-14T11:58:40.230 & OTTB & 284.20 & 15.58 & 5.05 & 9.75 & 85.89 & -10.25 & Yes & Yes \\
36 & 2022-10-14T12:47:44.045 & BB & 301.50 & 27.49 & 3.25 & 29.16 & 52.81 & -4.77 & No & Yes \\
37 & 2022-10-14T13:20:31.720 & PL & 291.35 & 24.31 & 7.06 & 8.90 & 244.57 & -44.63 & Yes & Yes \\
38 & 2022-10-14T13:21:36.375 & CPL & 288.70 & 23.05 & 2.93 & 9.55 & 691.62 & -16.21 & Yes & Yes \\
39 & 2022-10-14T13:38:25.400 & OTTB & 311.32 & 12.79 & 10.29 & 8.01 & 55.88 & -67.71 & Yes & Yes \\
40 & 2022-10-14T17:35:01.675 & SoftBand & 278.66 & 31.91 & 9.25 & 7.67 & 47.79 & -38.67 & No & Yes \\
41 & 2022-10-14T17:47:59.175 & SoftBand & 280.58 & 31.65 & 1.46 & 7.20 & 486.30 & -130.42 & No & Yes \\
42 & 2022-10-14T19:21:39.020 & OTTB & 302.72 & 18.95 & 5.94 & 8.03 & 321.19 & -10.00 & No & Yes \\
43 & 2022-10-14T20:48:47.425 & SoftBand & 286.06 & 37.31 & 11.61 & 6.97 & 254.97 & -102.61 & No & Yes \\
44 & 2022-10-15T02:13:54.545 & SoftBand & 288.25 & 4.63 & 1.26 & 11.28 & 128.58 & -17.94 & Yes & Yes \\
45 & 2022-10-16T10:43:17.080 & BB & 282.62 & 19.36 & 8.50 & 8.81 & 22.04 & -3.94 & No & Yes \\
46 & 2022-10-17T04:43:27.720 & CPL & 276.95 & 9.82 & 7.64 & 7.45 & 16.37 & -10.85 & Yes & No \\
47 & 2022-10-17T14:17:40.650 & OTTB & 291.35 & 21.34 & 10.91 & 10.52 & 47.56 & -15.97 & Yes & No \\
48 & 2022-10-17T15:14:07.600 & PL & 280.53 & 12.30 & 9.83 & 6.61 & 268.73 & -221.31 & No & Yes \\
49 & 2022-10-17T15:53:30.340 & BB & 279.93 & 16.88 & 14.97 & 7.50 & 13.36 & -12.99 & Yes & No \\
50 & 2022-10-17T16:35:38.085 & CPL & 281.66 & 12.46 & 8.73 & 9.82 & 48.81 & -26.24 & Yes & Yes \\
51 & 2022-10-18T13:57:25.965 & SoftBand & 286.87 & 2.99 & 10.27 & 7.23 & 4.82 & -10.47 & Yes & Yes \\
52 & 2022-10-18T16:22:09.650 & CPL & 300.25 & 25.07 & 8.56 & 11.35 & 295.70 & -5.49 & Yes & No \\
53 & 2022-10-22T04:44:52.700 & CPL & 300.97 & 29.09 & 3.36 & 21.45 & 86.99 & -4.82 & No & Yes \\
54 & 2022-10-22T16:30:16.000 & PL & 301.02 & 40.42 & 4.85 & 9.18 & 109.45 & -134.88 & Yes & No \\
55 & 2022-10-25T18:42:31.610 & BB & 277.10 & 4.59 & 12.13 & 7.22 & 14.00 & -6.36 & No & No \\
56 & 2022-10-31T23:08:59.180 & CPL & 277.14 & 15.63 & 2.91 & 10.49 & 14.27 & -4.90 & Yes & Yes \\
57 & 2022-11-09T09:05:47.515 & CPL & 291.45 & 14.72 & 3.44 & 8.74 & 298.00 & -10.95 & No & No \\
58 & 2022-11-09T14:33:10.135 & CPL & 288.10 & 19.74 & 3.98 & 26.27 & 432.66 & -7.65 & Yes & Yes \\
59 & 2022-11-09T15:29:02.435 & CPL & 286.99 & 18.24 & 3.06 & 8.52 & 502.87 & -1.47 & Yes & No \\
60 & 2022-11-09T15:58:26.975 & OTTB & 288.62 & 26.59 & 15.04 & 7.32 & 71.15 & -69.01 & Yes & Yes \\
61 & 2022-11-09T16:06:08.595 & CPL & 290.43 & 21.32 & 2.45 & 6.07 & 2537.76 & -3.71 & Yes & Yes \\
62 & 2022-11-09T16:49:33.240 & CPL & 294.07 & 32.15 & 10.32 & 14.52 & 643.86 & -14.48 & No & Yes \\
63 & 2022-11-09T16:50:28.385 & OTTB & 292.75 & 26.98 & 18.72 & 7.35 & 39.40 & -27.64 & No & Yes \\
64 & 2022-11-09T16:51:37.330 & PL & 280.08 & 26.93 & 8.11 & 13.33 & 31.15 & -5.87 & No & No \\
65 & 2022-11-09T16:52:01.750 & OTTB & 295.82 & 27.61 & 5.45 & 6.92 & 33.43 & 223.88 & No & No \\
66 & 2022-11-09T16:52:26.585 & OTTB & 278.83 & 23.98 & 9.29 & 6.68 & 111.64 & -29.51 & No & No \\
67 & 2022-11-09T16:54:17.565 & CPL & 295.70 & 23.93 & 3.80 & 15.72 & 359.39 & -13.91 & No & No \\
68 & 2022-11-09T16:55:20.665 & BB & 287.46 & 29.18 & 13.97 & 7.39 & 30.20 & -10.96 & No & No \\
69 & 2022-11-09T16:57:30.950 & BB & 278.25 & 15.69 & 17.94 & 7.45 & 69.34 & -13.18 & No & No \\
70 & 2022-11-09T16:59:56.580 & CPL & 290.30 & 20.83 & 3.57 & 19.85 & 100.90 & -3.42 & No & No \\
71 & 2022-11-09T17:05:04.670 & CPL & 286.18 & 22.81 & 4.97 & 12.97 & 480.98 & -1.80 & Yes & No \\
72 & 2022-11-09T17:21:41.635 & OTTB & 293.34 & 21.22 & 2.71 & 11.74 & 169.43 & -8.33 & Yes & No \\
73 & 2022-11-09T17:35:49.100 & OTTB & 303.74 & 24.36 & 4.51 & 10.84 & 331.01 & -29.99 & Yes & Yes \\
74 & 2022-11-09T17:46:18.390 & BB & 299.15 & 21.10 & 3.62 & 7.27 & 144.09 & 0.58 & Yes & Yes \\
75 & 2022-11-09T17:47:59.245 & CPL & 290.50 & 21.31 & 2.98 & 13.92 & 307.82 & -7.27 & Yes & Yes \\
76 & 2022-11-09T18:25:56.375 & CPL & 289.59 & 26.79 & 2.63 & 17.78 & 96.94 & -10.60 & No & Yes \\
77 & 2022-11-09T19:07:34.000 & CPL & 293.20 & 13.03 & 16.49 & 8.16 & 22.47 & -9.50 & Yes & Yes \\
78 & 2022-11-09T19:08:04.250 & SoftBand & 276.47 & 9.37 & 2.08 & 8.84 & 97.96 & -51.83 & Yes & Yes \\
79 & 2022-11-09T20:09:32.965 & BB & 289.05 & 27.24 & 6.08 & 10.98 & 69.20 & -15.85 & No & No \\
80 & 2022-11-09T20:45:50.955 & OTTB & 284.97 & 21.56 & 4.86 & 8.45 & 223.92 & -12.21 & Yes & Yes \\
81 & 2022-11-10T09:03:33.235 & CPL & 302.00 & 22.68 & 6.94 & 10.81 & 99.04 & -11.88 & Yes & No \\
82 & 2022-11-10T14:01:58.135 & OTTB & 284.30 & 13.73 & 7.35 & 6.67 & 25.53 & -5.10 & Yes & No \\
83 & 2022-11-10T18:10:56.205 & OTTB & 289.45 & 23.58 & 5.15 & 19.23 & 40.50 & -5.87 & No & Yes \\
84 & 2022-11-10T23:43:45.995 & CPL & 285.79 & 13.09 & 4.22 & 12.58 & 43.11 & -6.08 & Yes & No \\
85 & 2022-11-13T02:56:52.190 & CPL & 279.47 & 13.13 & 4.74 & 13.81 & 41.79 & -7.96 & No & No \\
86 & 2022-11-14T11:09:19.010 & BB & 285.86 & 14.33 & 5.62 & 13.49 & 53.19 & -4.31 & No & No \\
87 & 2022-11-15T17:14:34.865 & CPL & 294.84 & 21.22 & 2.91 & 10.28 & 181.28 & -10.70 & No & Yes \\
88 & 2022-11-19T18:42:35.280 & OTTB & 284.95 & 12.68 & 1.96 & 15.21 & 6.65 & -4.78 & Yes & Yes \\
89 & 2022-11-23T09:35:02.435 & PL & 294.73 & 7.75 & 14.95 & 7.37 & 4.97 & -5.21 & Yes & Yes \\
90 & 2022-11-23T19:17:25.100 & PL & 311.40 & 24.56 & 8.94 & 6.53 & 2.82 & -17.85 & Yes & Yes \\
91 & 2022-11-27T00:38:15.935 & PL & 310.60 & 35.64 & 10.30 & 7.98 & 8.41 & -8.30 & Yes & Yes \\
92 & 2022-11-30T01:05:01.125 & PL & 299.69 & 17.49 & 4.36 & 7.01 & 8.99 & -13.51 & No & Yes \\
93 & 2022-12-01T17:31:06.735 & BB & 276.56 & 38.99 & 4.63 & 7.40 & 9.78 & -9.50 & Yes & Yes \\
94 & 2022-12-05T16:51:36.925 & PL & 306.92 & 13.65 & 7.23 & 6.94 & 1.94 & -10.05 & Yes & No \\
95 & 2022-12-19T08:49:13.535 & BB & 284.86 & 20.70 & 17.26 & 7.52 & 6.97 & -7.84 & Yes & No \\
96 & 2022-12-20T13:09:17.025 & PL & 310.28 & 18.36 & 8.90 & 7.38 & 6.70 & -9.98 & Yes & Yes \\
97 & 2022-12-21T03:26:06.710 & CPL & 289.30 & 12.99 & 4.84 & 7.96 & 8.65 & -8.95 & Yes & Yes \\
\enddata
\tablenotetext{a}{Burst duration derived by the Bayesian blocks algorithm.}
\tablenotetext{b}{Burst start time relative to the trigger time using the Bayesian blocks algorithm.}
\end{deluxetable*}

\begin{figure*}
\label{fig:lc_example}
\caption{Representative examples of GECAM-detected bursts of SGR J1935+2154. For each burst, light curves are shown for high gain (HG, 6–300 keV). The background of each burst is shown by the red dotted line. The green vertical dotted line and blue dotted line are assessed by the Bayesian block (BB) algorithm. }
\epsscale{0.38}
\plotone{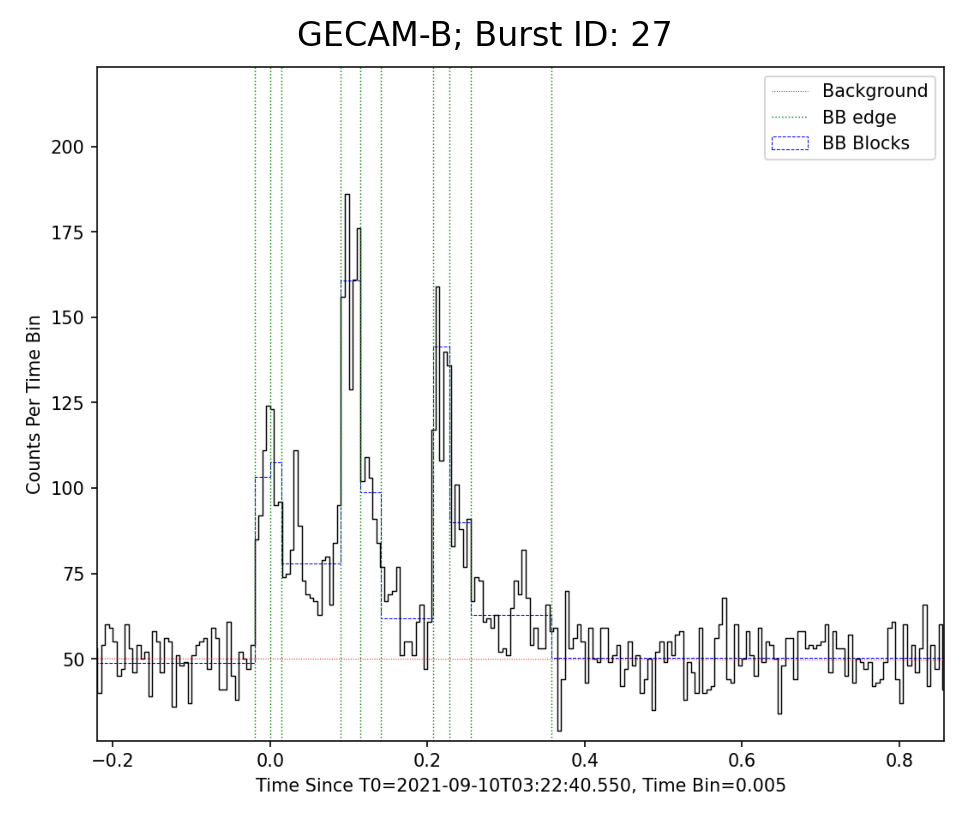}
\plotone{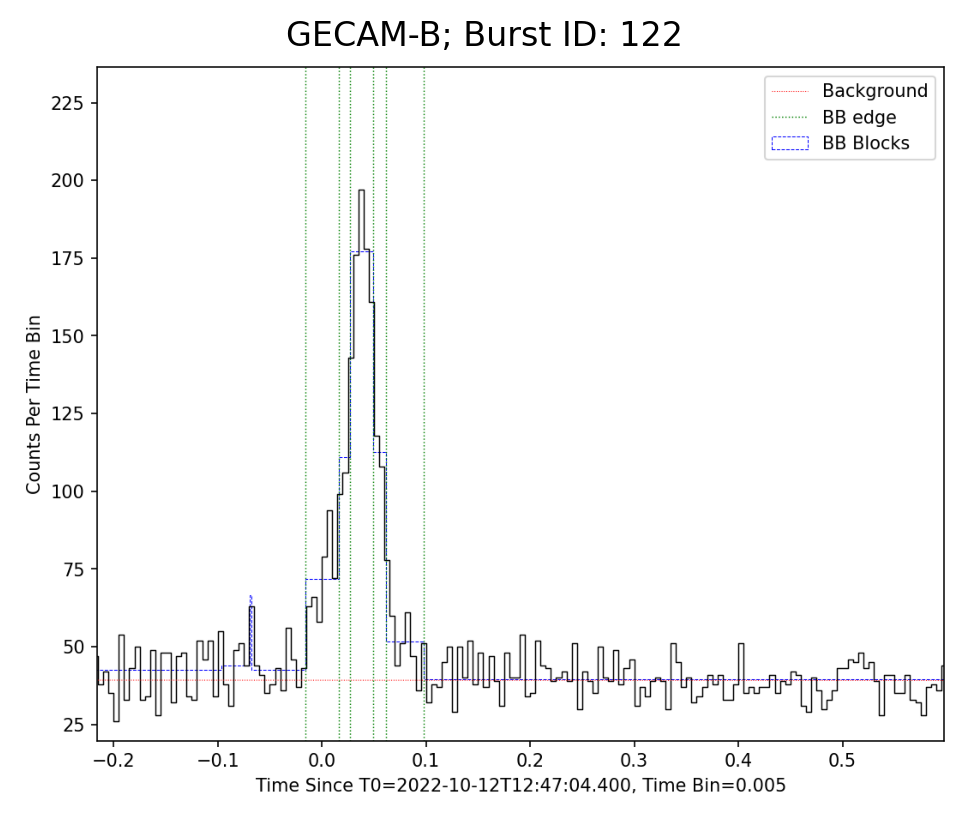}
\plotone{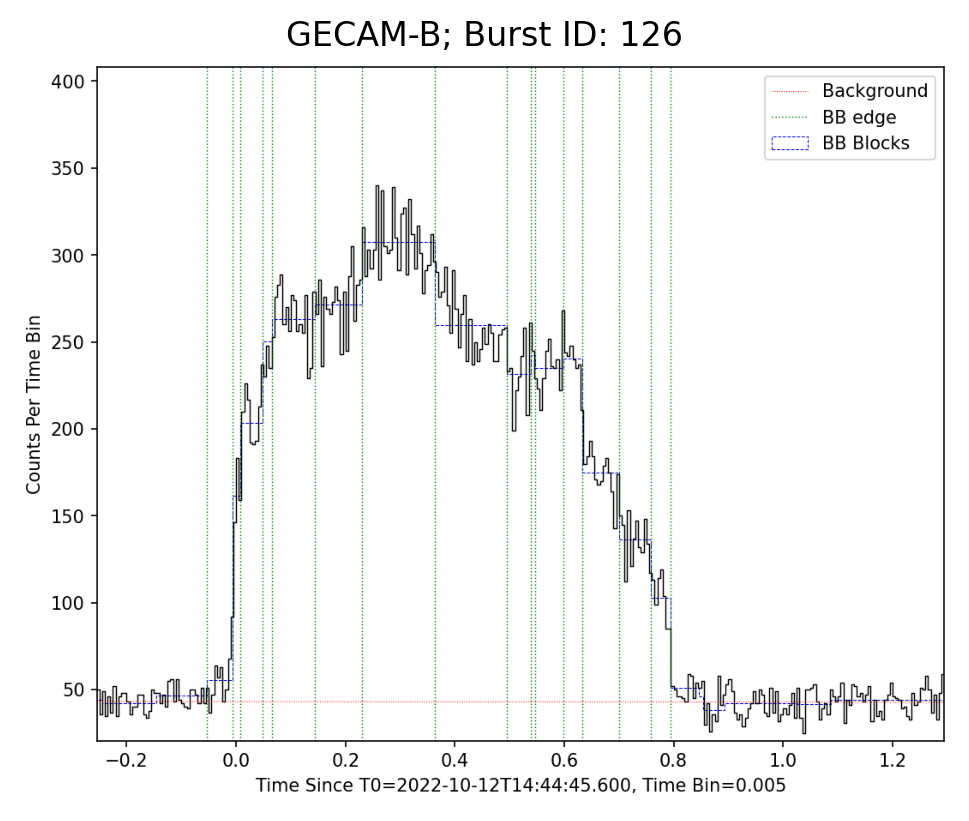}
\plotone{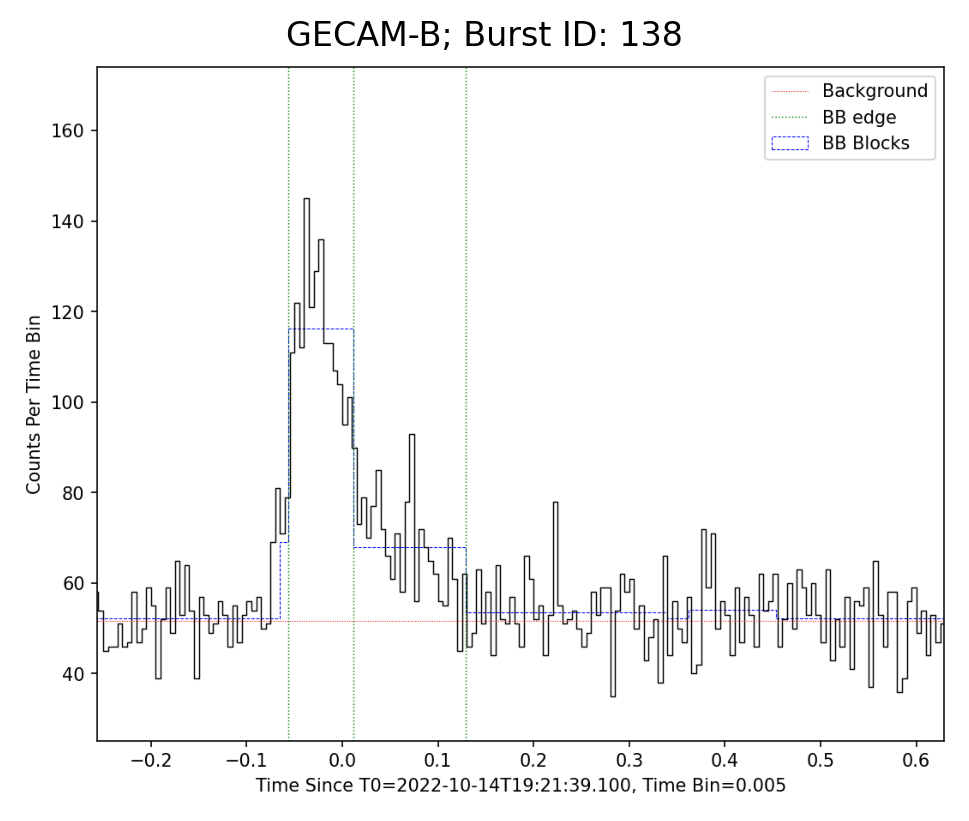}
\plotone{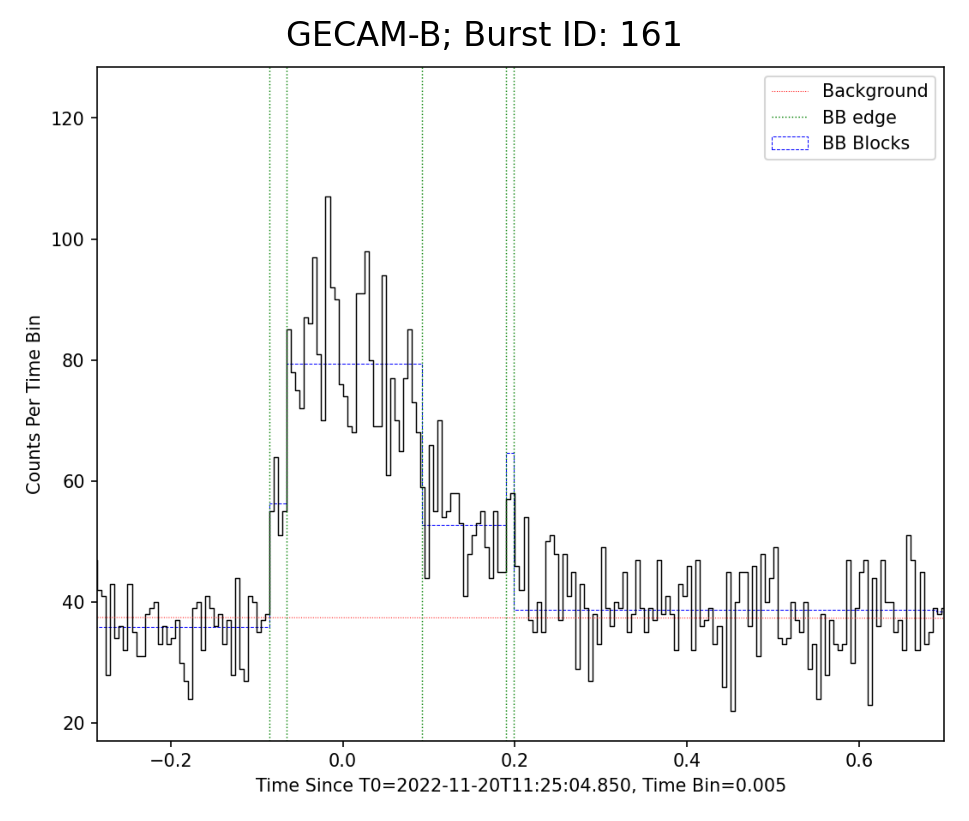}
\plotone{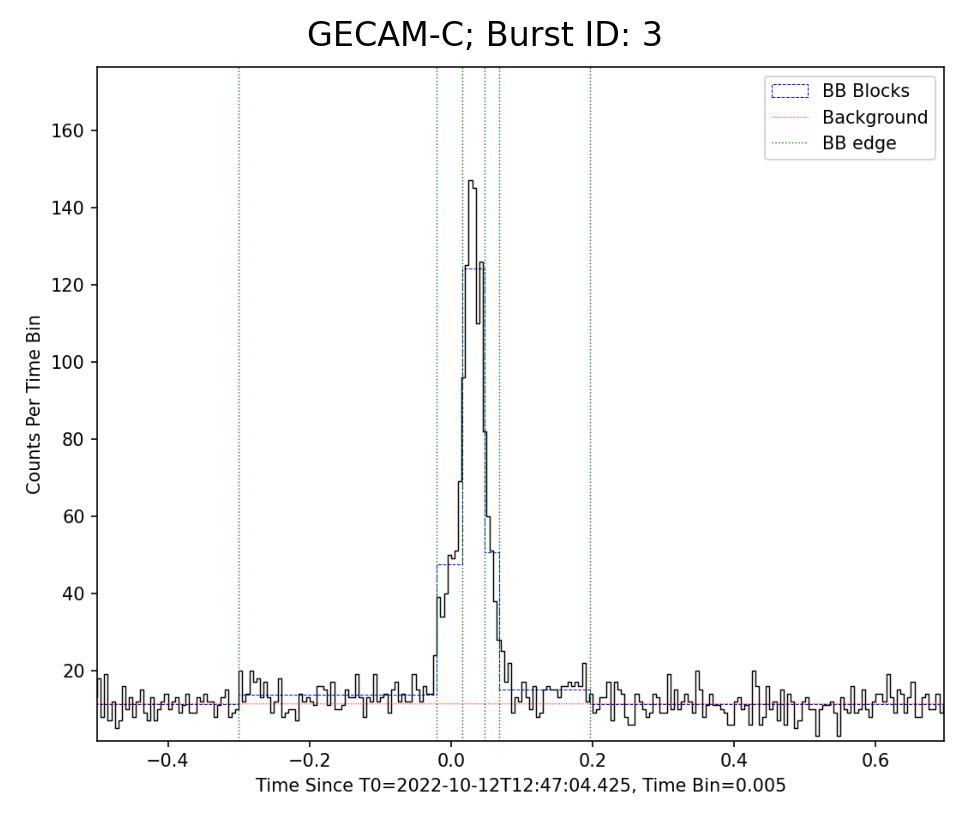}
\plotone{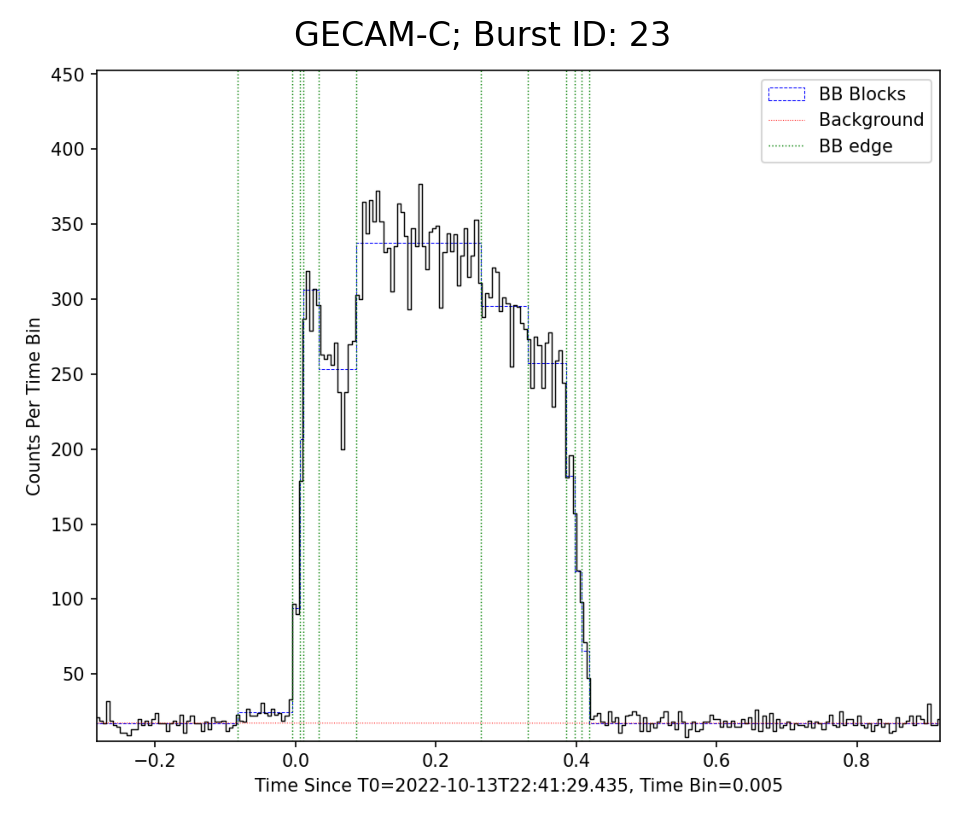}
\plotone{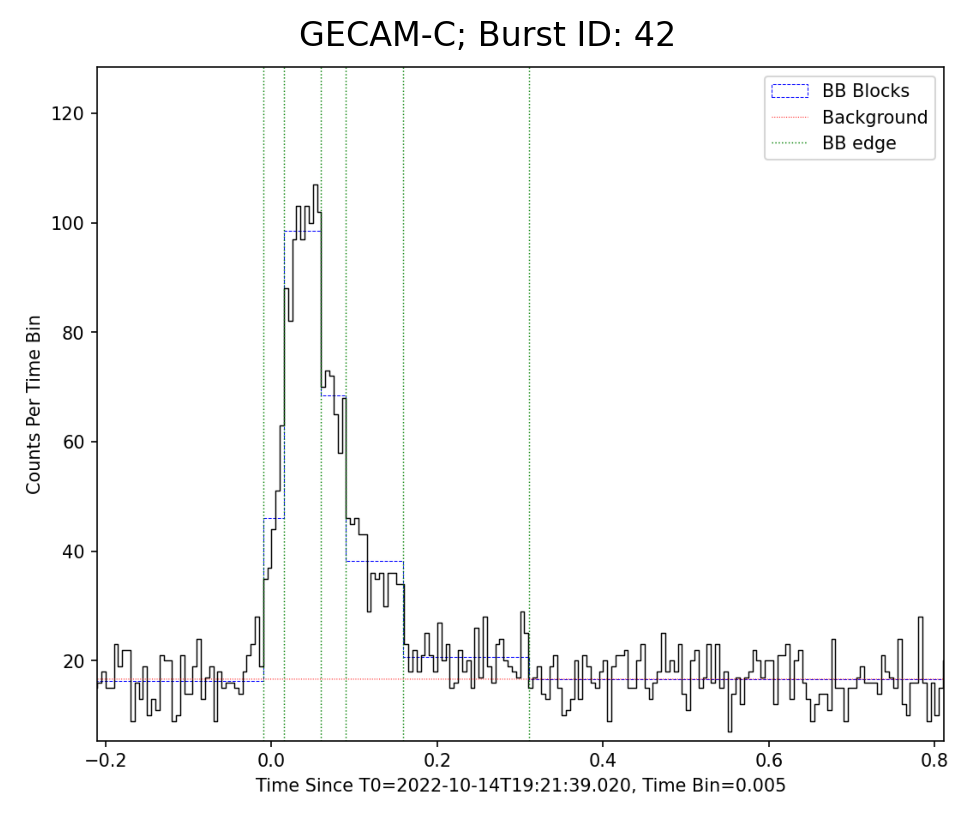}
\plotone{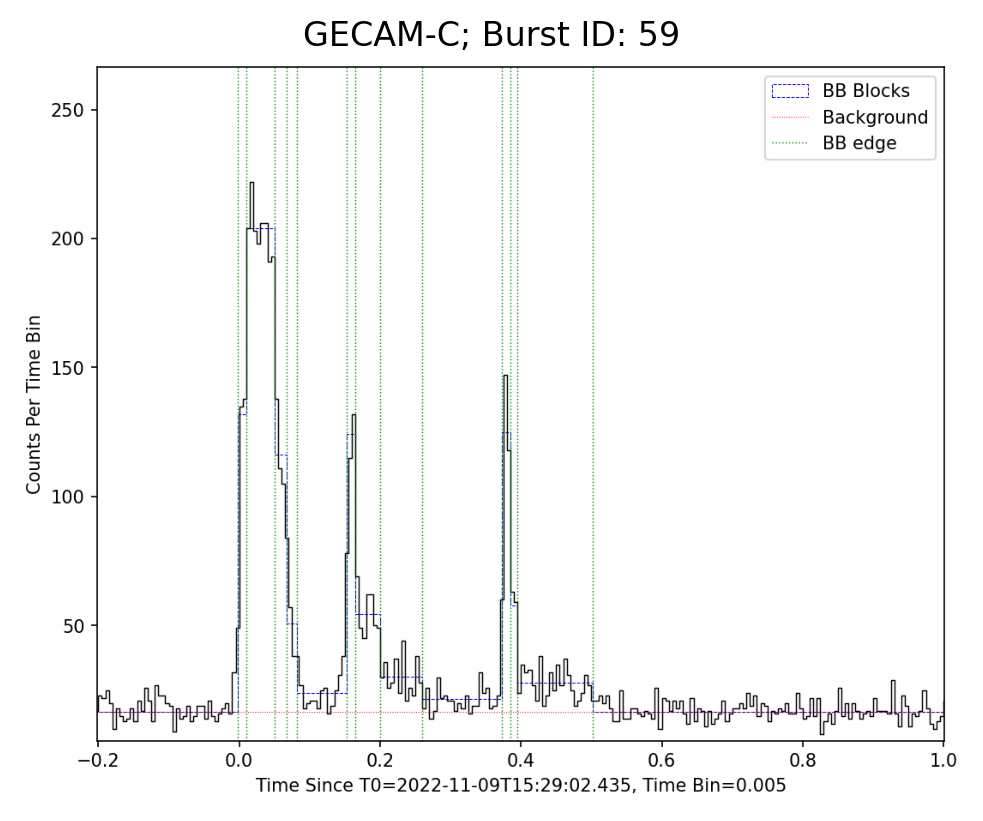}
\end{figure*}

\section{Catalog Analysis} \label{sec:cata_ana}
As shown above, we show light curves of 9 bursts in Fig. \ref{fig:lc_example} to display the variety of these bursts. The trigger time of each burst is listed in Tables \ref{tab:burst_list_gb} and \ref{tab:burst_list_gc}.
SGR J1935+2154 bursts consist of a single peak (such as ID: 122, 138, and 161 of GECAM-B; ID: 3 and 42 of GECAM-C), multiple peaks (ID: 27 of GECAM-B; ID: 59 of GECAM-C), and intermediate bursts (ID: 126 of GECAM-B; ID: 23 of GECAM-C).
The X-ray burst associated with the radio burst FRB 20221014 is ID: 138 of GECAM-B and ID: 42 of GECAM-C.

\subsection{Burst Activity}
The daily burst rate is defined as $N/P$, where $N$ is the observed burst number per day and $P$ represents the percentage of the effective observation time in a full day, excluding the time intervals when the Earth blocked SGR J1935+2154, or when GECAM passes through the South Atlantic Anomaly (SAA) area, during the instrument's startup period. 

\begin{figure*}
\epsscale{1.2}
\plotone{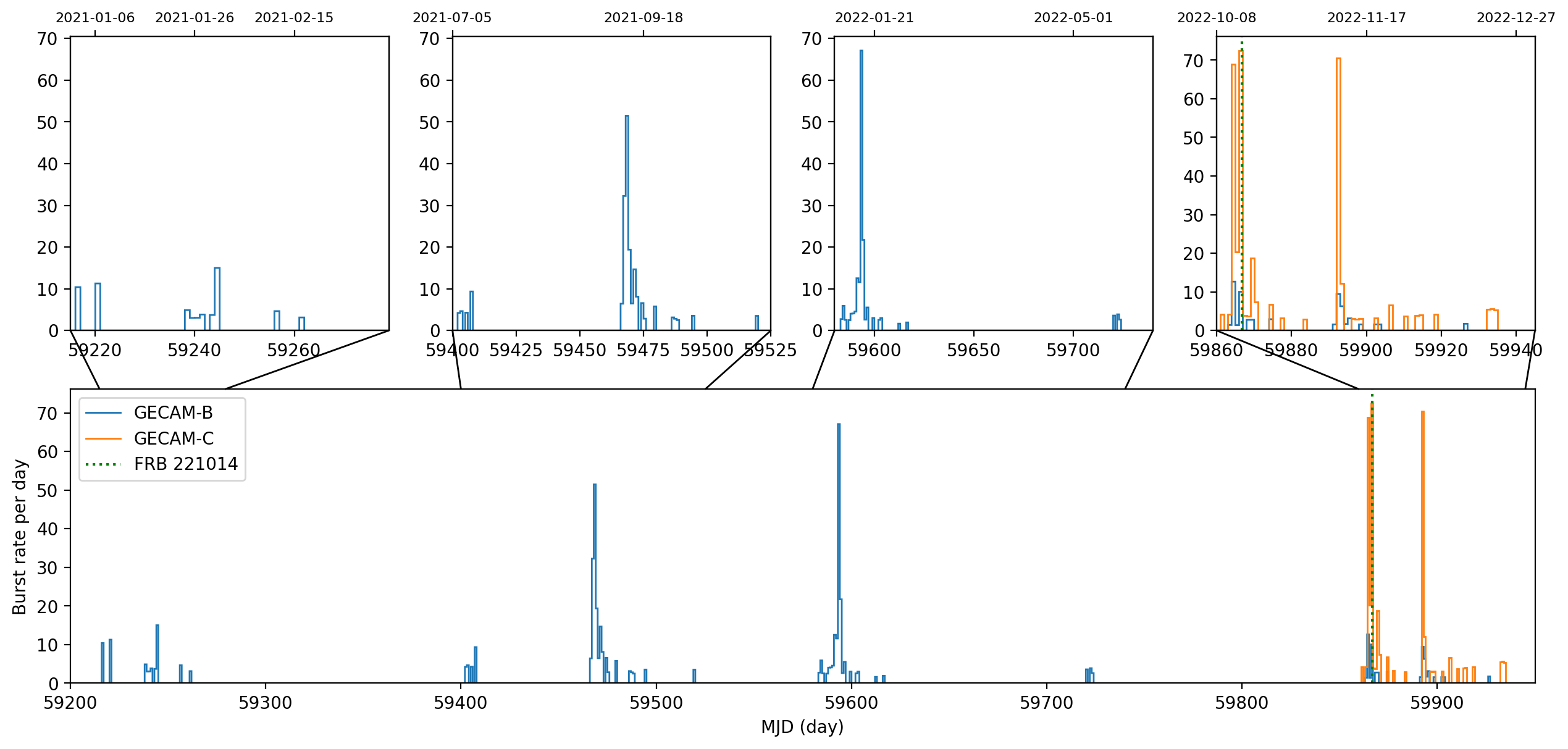}
\caption{The burst history of SGR J1935+2154 observed by GECAM-B/C.}
\label{fig:burst_history}
\end{figure*}

We calculate the effective observation time\footnote{GECAM-B orbits at low latitudes ($<30^\circ N/S$), while GECAM-C could cover nearly $90^\circ N/S$. We only calculate the observation time of GECAM-C at low latitudes in this paper since the background undergoes too violent changes to detect magnetar bursts when GECAM-C is across high latitudes.}
of GECAM-B/C individually and estimate the daily burst rate of GECAM-B/C, respectively (see Fig. \ref{fig:burst_history}).
Because the effective detectable energy range of GECAM-B changed in the HG channel ($>$40 keV after October 2022), weaker bursts may not be detected compared to GECAM-C observations.
Hence, given that the burst activity of GECAM-B is lower than that of GECAM-C ($>$15 keV), the daily burst rate estimated using GECAM-C data may be more reliable for assessing burst activity compared to using GECAM-B data after October 2022.
A detailed study on the evolution of the effective detectable energy range change of GECAM-B will be reported in Paper II.

\subsection{Period Search}
\begin{figure}
\epsscale{1.0}
\plotone{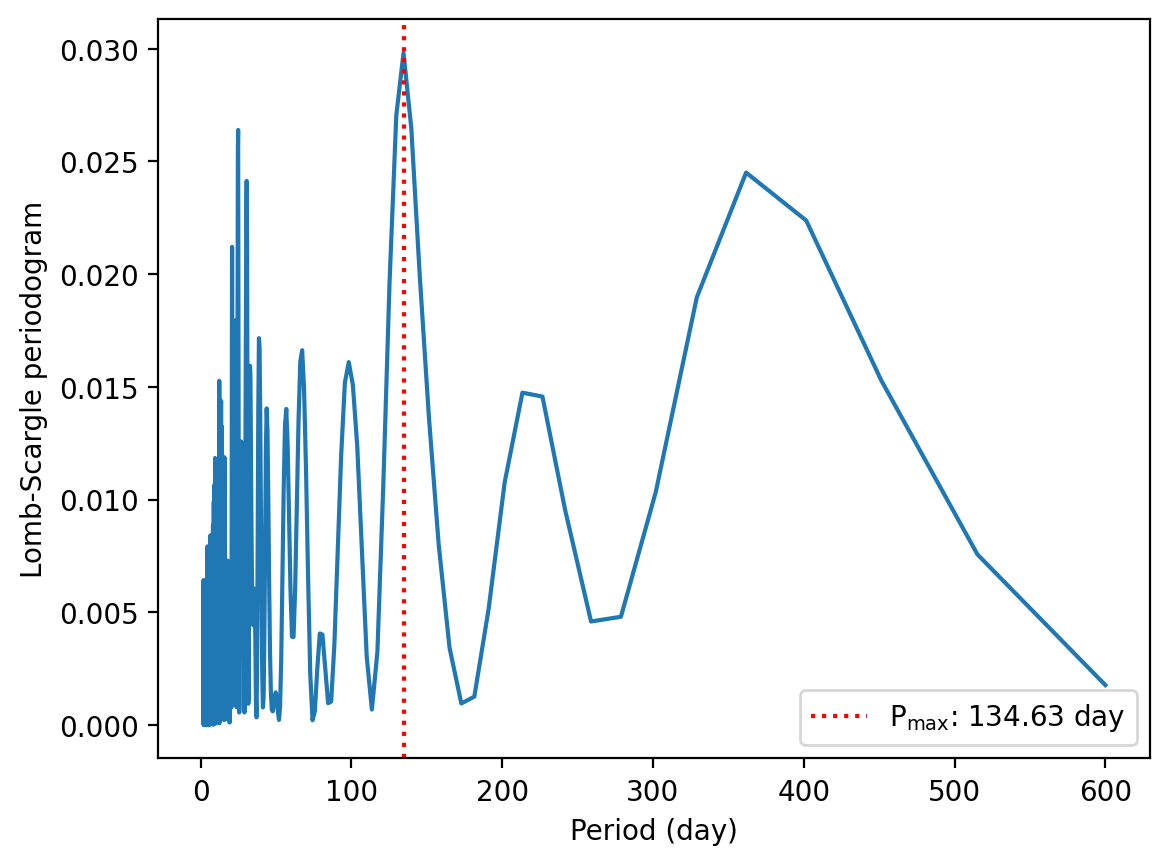}
\caption{Lomb-Scargle periodogram of X-ray bursts from SGR J1935+2154 observed by GECAM. The vertical red dotted line indicate the peak of periodogram which is 134.63 day.}
\label{fig:lomb_scargle}
\end{figure}

To explore the potential periodicity behavior in the burst history of SGR J1935+2154, we use the daily burst rate data of GECAM-B (before October 2022) and GECAM-C (after October 2022) to search periodicity by the Lomb-Scargle Periodogram method \citep{Lomb1976apss,Scargle1982apj,VanderPlas2018apjs}.
The periodogram results are shown in Fig \ref{fig:lomb_scargle}. The most significant peak of the Lomb-Scargle periodogram is 134.63 day (false alarm probability: 0.001, significance level $\sim 3 \sigma$). The peaks around the 50 day are likely caused by observation windows or bursts gap \citep{Xie2022mnras}.

\begin{figure}
\epsscale{1.0}
\plotone{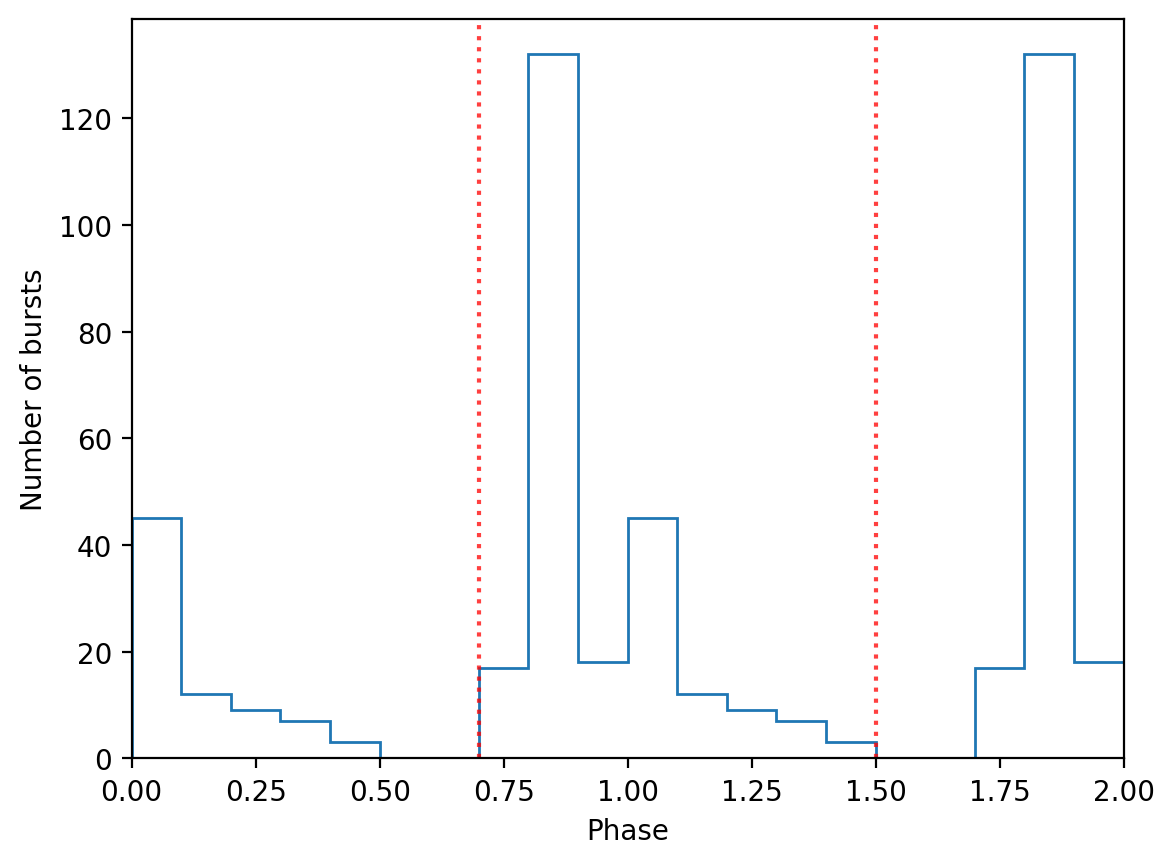}
\caption{Phase-folded burst rate according to the 134.63 day period with MJD 59215 (2021-01-01) referenced as phase 0. The active window (i.e. from $\phi=0.7$ to $\phi=1.5$) is denoted with red dotted lines.}
\label{fig:burst_phase}
\end{figure}

To uncover the burst activity of SGR J1935+2154 within a cycle, the start time of all bursts is folded into different phases at a given period $P=134.63$ day through the period folding method \citep{CHIMEFRB2020a,Zhang2021apj,Xie2022mnras},
\begin{equation}
    \phi=\frac{T-T_0}P-\mathrm{floor}\biggl(\frac{T-T_0}P\biggr),
\end{equation}
where $\phi$ is the folded phase, $T$ is the burst start time, $T_0$ is the phase start point (MJD: 59215; UTC: 2021-01-01), and the floor is a function that returns the floor of the input number. As shown in Fig \ref{fig:burst_phase}, the active day within a cycle (duty cycle) of SGR J1935+2154 bursts is 80\%.
The error of the period $P=134.63$ could be estimated to be 20 days with \citep{ChimeFrb2020Natur},
\begin{equation}
    \sigma = \frac{PW_{\mathrm{active}}}{T_{\mathrm{span}}},
\end{equation}
where $W_{\mathrm{active}}$ is the active days in one cycle and $T_{\mathrm{span}}$ is the longest time separation between burst arrival times.
Accordingly, the most probable period of SGR J1935+2154 from January 2021 to December 2022 is $134.63\pm20$ days, consistent with previous work \citep[][evaluated from July 2014 to January 2022]{Xie2022mnras} and the burst history over these two years could be grouped into 4 active episodes (see Fig. \ref{fig:burst_history}).

\subsection{Burst Duration}
\begin{figure}
\epsscale{1.0}
\plotone{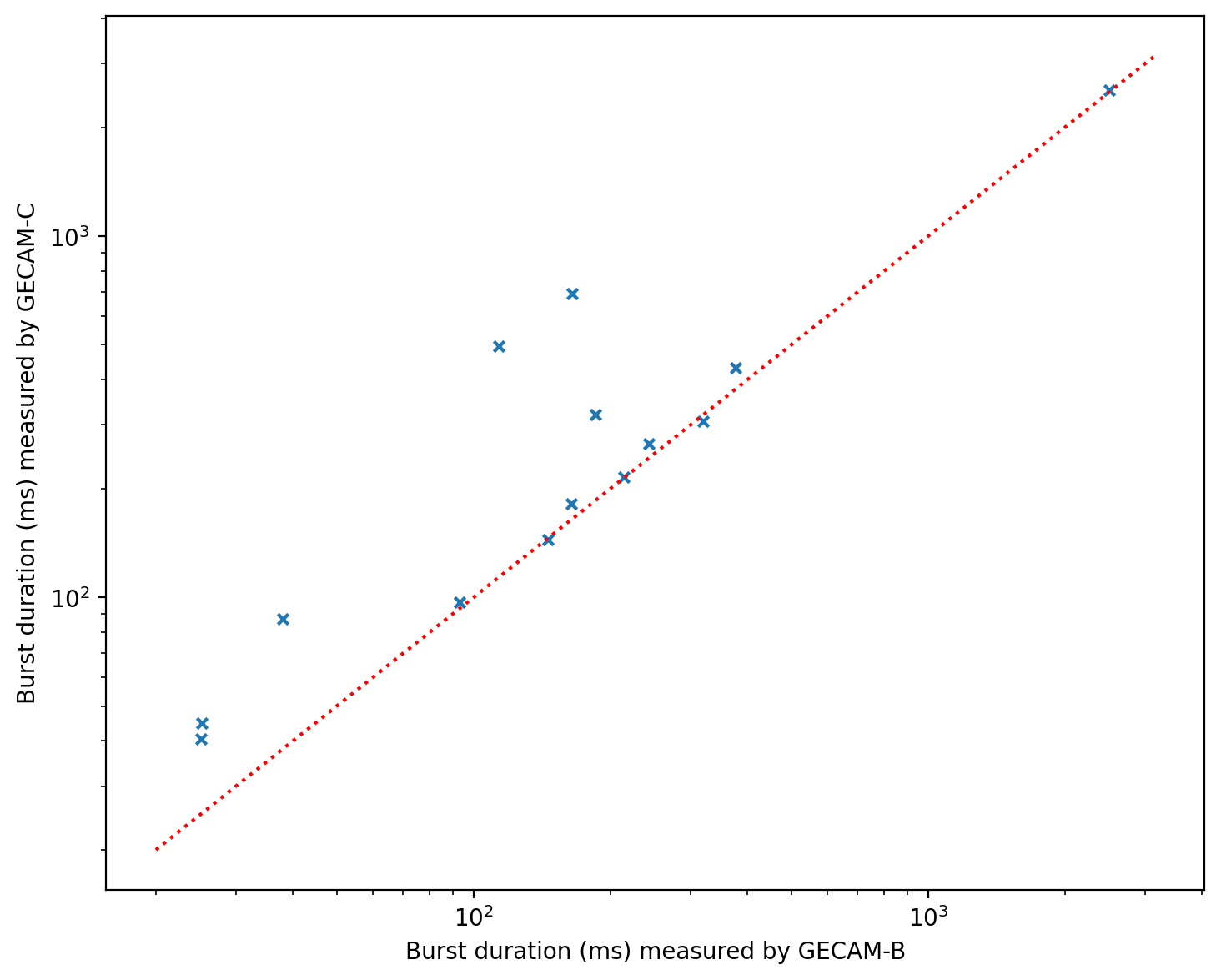}
\caption{The burst duration of 14 X-ray bursts observed by both GECAM-B and GECAM-C.}
\label{fig:burst_durat_B_C}
\end{figure}

As previous studies \citep{Lin2013ApJ}, the magnetar burst duration $T_\mathrm{bb}$ is calculated with the Bayesian blocks algorithm \citep{Scargle2013ApJ}. We measure the duration for each burst using the sliced event data of 10 s, including both pre- and post-burst trigger time intervals (from T$_0$-10 s to T$_0$+10 s, where T$_0$ is the burst trigger time). 
The blocks with a duration longer than 6 s (at least twice the spin period of SGR J1935+2154) are treated as background. The background count rate is estimated by fitting a first-order polynomial to the background blocks.
A block with a significance larger than $5\sigma$ is considered to be part of the burst region.
The first block with a significance higher than $5\sigma$ is marked as the burst start time (see Tables \ref{tab:burst_list_gb} and \ref{tab:burst_list_gc}).
As an example, the blocks and block edges of a burst are illustrated in Fig. \ref{fig:lc_example}.
However, some bursts with multiple peaks (e.g., burst \#59 of GECAM-C in Fig. \ref{fig:lc_example}) have at least two subsequent blocks, along with quiescent intervals. In such cases, the duration of the entire burst is calculated from the start of the first burst block to the end of the last block.
Due to the different effective detectable energy ranges between GECAM-B and GECAM-C, 14 bursts observed by both GECAM-B\&C exhibit different durations (see Fig. \ref{fig:burst_durat_B_C} or see an example of X-ray burst \#138 of GECAM-B and \#42 of GECAM-C in Fig. \ref{fig:lc_example} associated with FRB 221014).

\begin{figure*}
\epsscale{1.0}
\plotone{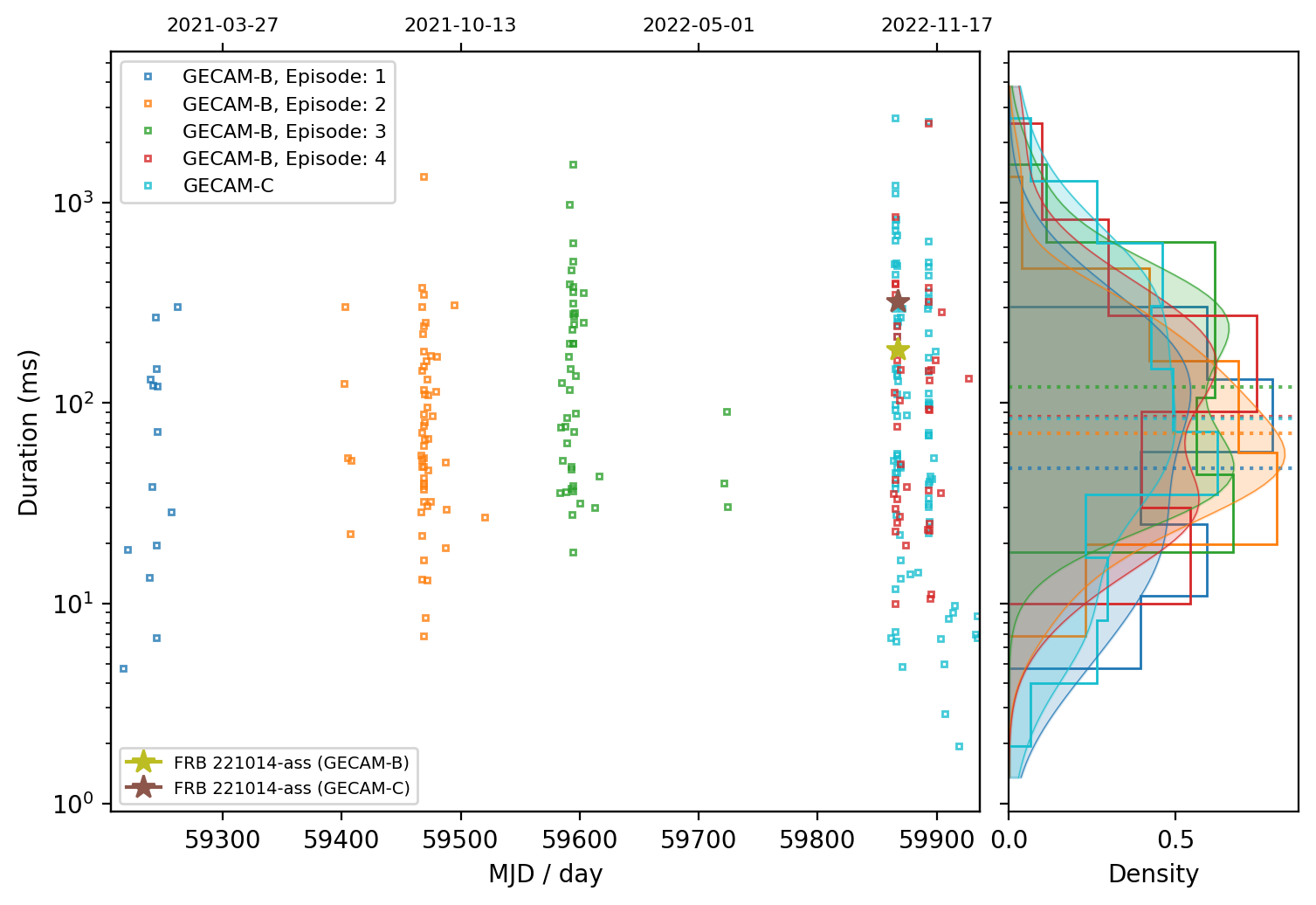}
\caption{The left panel is the burst duration of each burst observed by GECAM-B/C. The right panel is the burst duration distributions of GECAM-B/C. The horizontal dotted lines show the best-fit lognormal function with a mean value listed in Table \ref{tab:burst_durat}.}
\label{fig:dist_burst_durat}
\end{figure*}

The duration ($T_\mathrm{bb}$) of each burst observed by GECAM-B/C is listed in Tables \ref{tab:burst_list_gb} and \ref{tab:burst_list_gc}. The burst duration distributions of GECAM-B/C are shown in Fig. \ref{fig:dist_burst_durat}. These distributions are well fitted with lognormal functions, as illustrated in Table \ref{tab:burst_durat}.
We perform the one-sample Kolmogorov-Smirnov test to compare the distribution of GECAM-B/C against a lognormal distribution and show the goodness of fit in Table \ref{tab:burst_durat}.
The mean values of burst duration for the four active episodes are $47.34_{-34.35}^{+125.19}$ ms for the first episode, $70.21_{-45.05}^{+125.69}$ ms for the second episode, $120.75_{-79.32}^{+231.22}$ ms for the third episode, and $85.51_{-60.77}^{+210.03}$/$83.33_{-66.32}^{+324.99}$ ms (given by GECAM-B/C data) for the fourth active episodes. The full sample gives the mean value $83.48_{-61.57}^{+234.64}$ ms.
The results show that the distribution of burst duration detected by GECAM from SGR J1935+2154 exhibits a lognormal distribution, which is consistent with previous work \citep{Lin2020a,Lin2020b,Cai2022apjs_A,Rehan2023ApJ,Rehan2024ApJ}.

\movetableright=-1cm
\begin{table}
\footnotesize
\caption{The mean of burst duration distribution in each episode obtained by fitting to a lognormal function.}
\label{tab:burst_durat}
\begin{tabular}{ccccc}
\hline
Instrument & Episode & Mean & Statistic\tablenotemark{a} & p-value \\
\hline
GECAM-B & 1st episode & $47.34_{-34.35}^{+125.19}$ & 0.2 & 0.60 \\
        & 2nd episode & $70.21_{-45.05}^{+125.69}$ & 0.07 & 0.95 \\
        & 3rd episode & $120.75_{-79.32}^{+231.22}$ & 0.11 & 0.57 \\
        & 4th episode & $85.51_{-60.77}^{+210.03}$ & 0.13 & 0.48 \\
GECAM-C & 4th episode & $83.33_{-66.32}^{+324.99}$ & 0.07 & 0.69 \\
\hline
\end{tabular}
\tablenotetext{a}{one-sample Kolmogorov-Smirnov test statistic}
\end{table}

\subsection{Burst Waiting Time}
\begin{figure*}
\epsscale{1.0}
\plotone{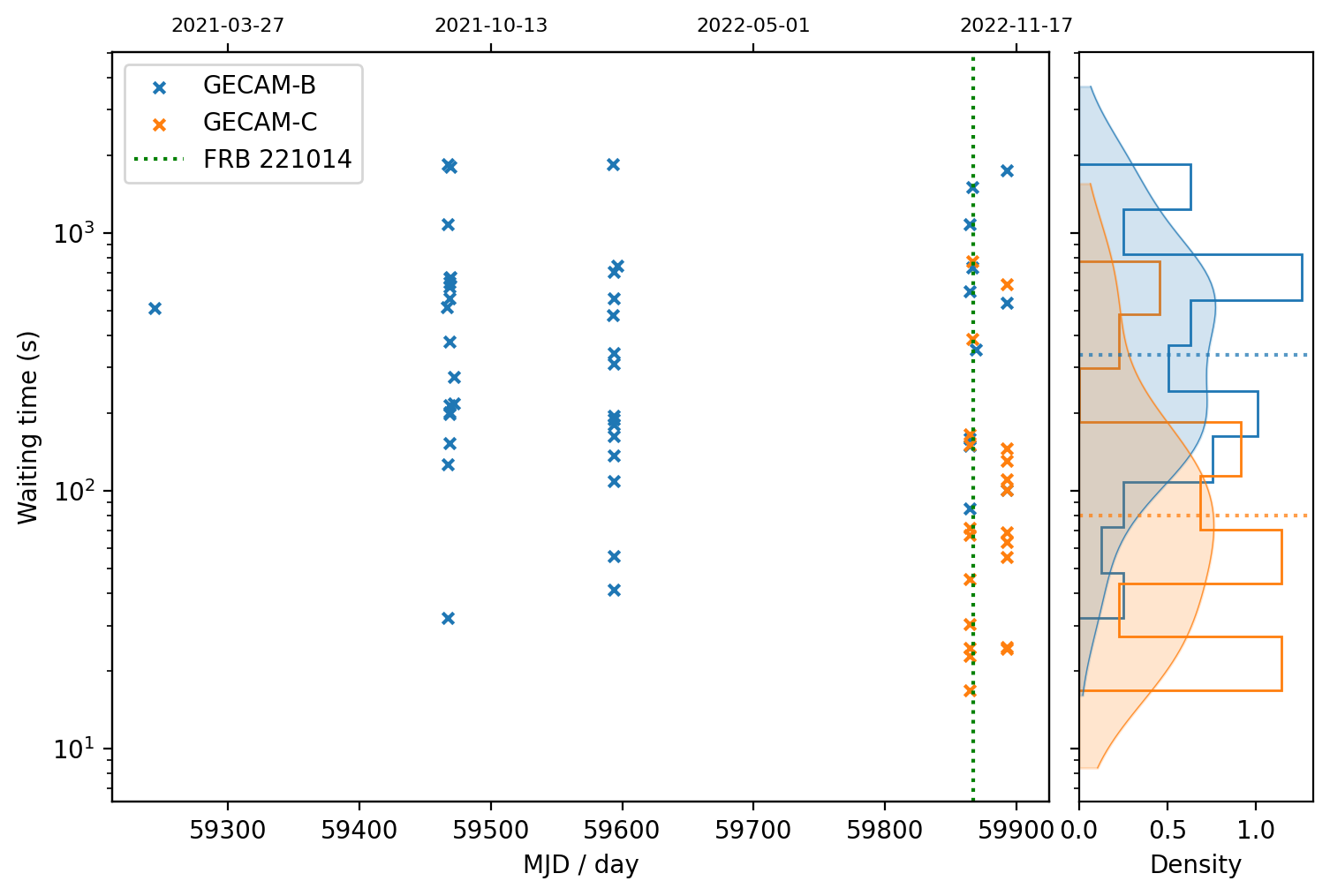}
\caption{The left panel is the waiting time (time interval between two successive bursts observed by GECAM-B/C) of the X-ray bursts v.s their occurring times, where the time of the FRB 20221014 is represented by the green dashed line.
The right panel is the distribution of waiting time. The horizontal dotted lines represent the best-fit lognormal function with a mean of $338.84_{-212.95}^{+573.16}$ seconds for GECAM-B and $79.43_{-51.89}^{+149.65}$ seconds for GECAM-C, respectively.}
\label{fig:wt_history}
\end{figure*}

The waiting time between the successive bursts that fall within an uninterrupted observation time interval is defined as
\begin{equation}
\Delta{t}=t_{\mathrm{i}}-t_{\mathrm{i-1}}
\end{equation}
where $t_{i}$ is the burst start time of the $i-$th burst.
As shown in Fig. \ref{fig:wt_history}, there are 45 waiting times of GECAM-B and 21 waiting times of GECAM-C during continuous observation time intervals.

Fig. \ref{fig:wt_history} shows the distribution of waiting time measured by GECAM-B/C.
These distributions could be well fitted with lognormal functions, resulting in a mean value of $338.84_{-212.95}^{+573.16}$ seconds for GECAM-B and a mean value of $79.43_{-51.89}^{+149.65}$ seconds for GECAM-C. Compared to a lognormal distribution, the Kolmogorov-Smirnov tests yield a statistic of 0.104 and a p-value of 0.672 for GECAM-B, and a statistic of 0.112 and a p-value of 0.929 for GECAM-C.
This lognormal distribution behavior is similar to that observed in previous studies \citep{Cai2022apjs_A,Xie2024ApJ}.
Since the relatively short continuous observation time intervals of GECAM-C for observing SGR J1935+2154 (only dozens of minutes) compared to GECAM-B, it is difficult to measure the longer waiting times between more successive bursts. Therefore, the distribution's mean value of GECAM-C is smaller than that of GECAM-B.

\subsection{Burst Hardness Ratio}
The net counts of the source is estimated as
\begin{equation}
    N_\mathrm{i}=C_\mathrm{i}-B_\mathrm{i},
\end{equation}
where $C_\mathrm{i}$ and $B_\mathrm{i}$ represent the total counts and background counts in the burst duration, respectively. The hardness ratio is the ratio of net counts in different energy ranges.

\begin{figure}
\epsscale{1.0}
\plotone{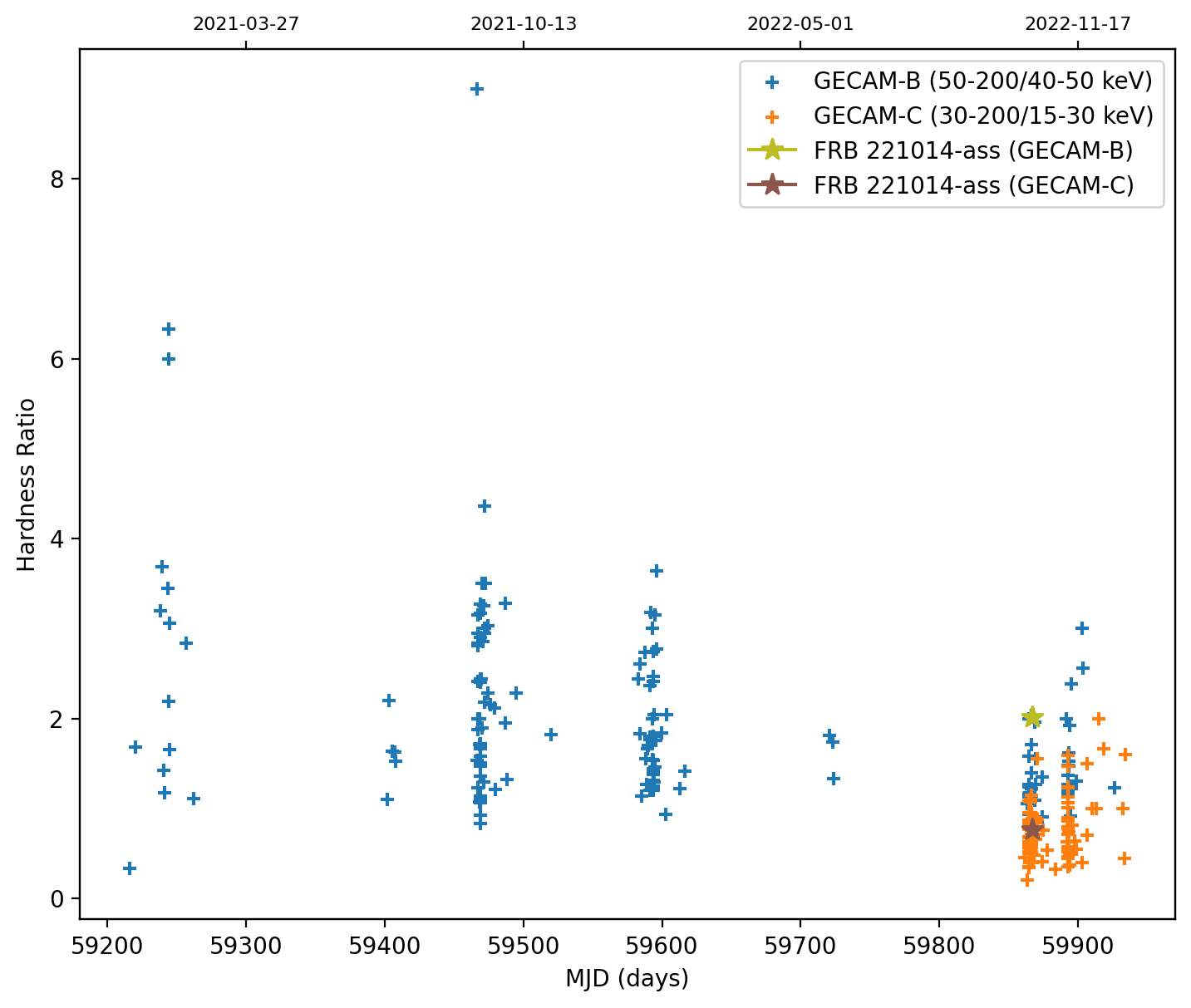}
\caption{The evolution of hardness ratio of each burst in 50-200/40-50 keV for GECAM-B and 30-200/15-30 keV for GECAM-C.}
\label{fig:hard_ratio}
\end{figure}

Since changes in the effective detectable energy range of GECAM-B over the two years, we measure the hardness ratio of all bursts within the same effective range of 50-200/40-50 keV for GECAM-B. The hardness ratio of GECAM-C is calculated in the range of 30-200/15-30 keV.
The hardness ratio of the four active episodes for the GECAM-B data yields mean values of $2.72\pm1.70$, $2.25\pm1.20$, $1.87\pm0.63$, and $1.48\pm0.47$, respectively.
The hardness ratio of the last episode for the GECAM-C data yields a mean value of $0.76\pm0.33$.
It presents that the evolution of the burst hardness tends to be softer through the observation time, as shown in Fig. \ref{fig:hard_ratio}. More detailed spectral properties and evolution will be reported in Paper II.

\section{Summary and Discussions} \label{sec:summa_discus}
In this paper, we perform a targeted sub-threshold search GECAM daily EVT data for SGR J1935+2154 bursts during the year of 2021 and 2022. We design five spectral templates to target search and find 159 X-ray bursts from GECAM-B and 97 bursts from GECAM-C (refer to Tables \ref{tab:burst_list_gb} and \ref{tab:burst_list_gc}).
Among the templates, CPL, OTTB, and BB cover a higher proportion of them compared to others (refer to Table \ref{tab:temp_frac}), providing more accuracy in location (see Fig. \ref{fig:dist_ang_sep}). Therefore, a simple power-law function and exponential cutoff function with softer parameter settings might be preferred for searching magnetar X-ray bursts.

The burst activity is evaluated by estimating the daily burst rate data, as illustrated in Fig. \ref{fig:burst_history}. We employ the Lomb-Scargle method to investigate potential periodic behavior, which results in the most likely period of $134.63\pm20$ days (refer to Fig. \ref{fig:lomb_scargle}). Therefore, the burst history over these two years could be divided into four active episodes (see Fig. \ref{fig:burst_history}).
The period of $134.63\pm20$ days with 80\% duty cycle is well consistent with previous study \citep[][evaluated from July 2014 to January 2022]{Xie2022mnras}.
Unlike other periods derived in previous work \citep[$\sim 231$ days evaluated from 2014 to 2020 or $\sim 238$ days evaluated from July 2014 to October 2021 in][respectively]{Grossan2021PASP,Zou2021apjl}, a shorter period is found in this study, possibly indicating an increased activity of X-ray bursts from SGR J1935+2154 after the year of 2021.

\begin{table}
\footnotesize
\caption{Comparison of the burst duration of SGR 1935+2154 burst episodes since its discovery in 2014, as Observed by the Fermi/GBM and GECAM-B/C.}
\label{tab:burst_durat_compar}
\begin{tabular}{cccc}
\hline
Episode & Instrument & Number of bursts & Duration\tablenotemark{a} \\
 & & & (ms) \\
\hline
Jul 2014 & GBM & 3 & ... \\
Feb-Mar 2015 & GBM & 24 & $78.00_{-14.00}^{+17.00}$ \\
May-Jun 2016 & GBM & 42 & $72.00_{-6.00}^{+7.00}$ \\
Jun-Jul 2016 & GBM & 54 & $128.00_{-10.00}^{+11.00}$ \\
Nov 2019 & GBM & 22 & $121.00_{-33.00}^{+45.00}$ \\
Apr-May 2020 & GBM & 151 & $182.00_{-19.00}^{+22.00}$ \\
Jan 2021 & GECAM-B & 14 & $47.34_{-34.35}^{+125.19}$ \\
Sep 2021 & GBM & 79 & $77.34\pm1.03$ \\
 & GECAM-B & 57 & $70.21_{-45.05}^{+125.69}$ \\
Jan 2022 & GBM & 112 & $97.03\pm1.03$ \\
 & GECAM-B & 46 & $120.75_{-79.32}^{+231.22}$ \\
Oct-Dec 2022 & GECAM-B & 42 & $85.51_{-60.77}^{+210.03}$ \\ 
 & GECAM-C & 97 & $83.33_{-66.32}^{+324.99}$ \\
\hline
\end{tabular}
\tablenotetext{a}{The lognormal mean value of burst duration.}
\tablecomments{results of Fermi/GBM observations are obtained from \cite{Lin2020a,Lin2020b,Rehan2023ApJ,Rehan2024ApJ}.}
\end{table}

We perform extensive studies on the statistical characteristics of these X-ray bursts, including the burst duration, waiting time, and hardness ratio.
The duration follows a lognormal distribution for GECAM-B/C, respectively (see Fig. \ref{fig:dist_burst_durat} and Table \ref{tab:burst_durat}). Due to the varying effective detectable energy ranges of GECAM-B and GECAM-C, some bursts detected by both satellites may show different durations (see Fig. \ref{fig:lc_example} for examples). In such cases, the duration measured by GECAM-C is more informative across the entire energy range when both satellites detect the same burst.
The burst duration of SGR J1935+2154 is around 100 ms, see Table \ref{tab:burst_durat_compar} for comparison among different burst episodes.
The waiting time also conforms to a lognormal function with mean values of $338.84_{-212.95}^{+573.16}$ seconds and $79.43_{-51.89}^{+149.65}$ seconds for GECAM-B/C, respectively (refer to Fig. \ref{fig:wt_history}). Because of the relatively short continuous observation time intervals for GECAM-C observing SGR J1935+2154 compared to GECAM-B, the distribution's mean value of GECAM-C is smaller than that of GECAM-B. Such lognormal distribution behavior is similar to that observed in previous studies \citep{Cai2022apjs_A,Xie2024ApJ}.
The hardness ratio of X-ray bursts tends to become softer over the course of the observation time spanning these two years (see Fig. \ref{fig:hard_ratio}).

These statistical results indicate that the X-ray bursts from SGR J1935+2154 show increased activity, as observed by Fermi/GBM \citep[e.g.,][]{Rehan2024ApJ}.
The magnetar bursts also exhibit a softer spectrum over the two-year period, as presented by the temporal evolution of the hardness ratio and a detailed spectral analysis will follow.
Similar to FRB 200428, which occurred during the most active episode (April-May 2020) preceding 2020, the radio burst FRB 20221014
is detected in the final and most active episode (October-December 2022).
In view of their millisecond duration and high energy releases, FRBs are widely suggested to originate from the violent activities of compact objects, in particular magnetars \citep{Popov2010,Kulkarni2014,Katz2016,Connor2016,Cordes2016,Lyutikov2017}.
The mechanisms leading to such burst activities may correspond to some sudden changes in the magnetic configuration of the magnetars \citep[e.g., starquakes, magnetic field reconnections][]{Thompson1995MNRAS,Thompson1998PhRvD,Jones2003ApJ,Levin2012MNRAS}.
In that case, FRB emission may result from such burst interactions (e.g., the collision between different Alfven waves or different explosion outflows) in a global activity of a magnetar and then a detected FRB be associated with an X-ray burst \citep{Yang2021ApJ,Xie2024ApJ}.

This work is supported by the Strategic Priority Research Program of the Chinese Academy of Sciences (Grant No. XDB0550300, XDA30050000), the National Key R\&D Program of China (2021YFA0718500), the National SKA program of China (2020SKA0120300), the National Natural Science Foundation of China (Grant No. 12393811, 12173038, 12303045, 12273042), the Strategic Priority Research Program on Space Science of the Chinese Academy of Sciences (Grant No. XDA15360102, XDA15360300, XDA15052700), the Natural Science Foundation of Hebei Province (No. A2023205020), and the Science Research Program of Dezhou University (2024xjrc142).



\bibliography{main}{}
\bibliographystyle{aasjournal}

\end{document}